\documentclass[journal]{IEEEtran}
\usepackage{amsfonts,epsfig,cite,array,graphicx,amsmath,amssymb,epstopdf}
\usepackage{caption, longtable, makecell}
\usepackage{multirow}
\usepackage{amsmath}
\usepackage{booktabs}
\usepackage{caption}
\usepackage{capt-of}
\usepackage{balance}
\usepackage{subfigure}
\usepackage{fixltx2e}
\usepackage{setspace}
\usepackage{color, colortbl}
\usepackage{adjustbox}
\definecolor{Gray1}{gray}{0.9}
\definecolor{LightCyan}{rgb}{0.95,1,1}
\definecolor{Lightyellow}{rgb}{0.98,0.92,0.92}
\definecolor{Lightpink}{rgb}{1,0.95,0.98}
\definecolor{Lightgreen}{rgb}{0.95,0.99,0.88}
\definecolor{Lightgray}{rgb}{0.98,0.98,0.90}
\definecolor{Lightaqya}{rgb}{0.95,0.86,0.80}
\definecolor{Lightplum}{rgb}{0.95,0.99,0.92}
\usepackage{mathtools}

\usepackage{algorithm}
\usepackage{algpseudocode}
\begin{document}

\title{BiCurNet: Pre-Movement EEG based Neural Decoder for Biceps Curl Trajectory Estimation}

\author{Manali Saini*, Anant Jain*, Suriya Prakash Muthukrishnan, Shubhendu Bhasin, Sitikantha Roy~and~Lalan Kumar
\thanks{*Manali Saini and Anant Jain have contributed equally to this work.}
\thanks{This work was supported in part by DRDO - JATC project with project number RP04191G.}
\thanks{This work involved human subjects or animals in its research. Approval of all ethical and experimental procedures and protocols was granted by the Institute Ethics Committee, All India Institute of Medical Sciences, New Delhi, India with reference number IEC-751/07.08.2020,RP-06/2020.}
\thanks{Manali Saini is with the Department of Electrical Engineering, Indian Institute of Technology Delhi, New Delhi 110016, India (e-mail: manaliigit@gmail.com).}
\thanks{Anant Jain is with the Department of Electrical Engineering, Indian Institute of Technology Delhi, New Delhi 110016, India (e-mail: anantjain@ee.iitd.ac.in).}
\thanks{Suriya Prakash Muthukrishnan is with the Department of Physiology, All India Institute of Medical Sciences, New Delhi - 110016, India(e-mail: dr.suriyaprakash@aiims.edu).}
\thanks{Subhendu Bhasin is with the Department of Electrical Engineering, Indian Institute of Technology Delhi, New Delhi 110016, India (e-mail: sbhasin@ee.iitd.ac.in).}
\thanks{Sitikantha Roy is with the Department of Applied Mechanics, Indian Institute of Technology Delhi, New Delhi 110016, India (e-mail: sroy@am.iitd.ac.in).}
\thanks{Lalan Kumar is with the Department of Electrical Engineering, Bharti School of Telecommunication, and Yardi School of Artificial Intelligence, Indian Institute of Technology Delhi, New Delhi 110016, India (e-mail: lkumar@ee.iitd.ac.in).}
}
\maketitle
\begin{abstract}
Kinematic parameter (KP) estimation from early electroencephalogram (EEG) signals is essential for positive augmentation using wearable robots. However, surface EEG-based early KP estimation studies are sparse in the literature. In this study, simultaneous surface EEG and kinematics data of five participants is collected during the biceps-curl motor task. The feasibility of early estimation of KPs is demonstrated using brain source imaging (BSI). Discrete wavelet transform (DWT) is utilized for sub-band extraction from pre-processed EEG signals. Further, spherical and head harmonics domain features are extracted from sub-bands of the EEG signals. A deep-learning based decoding model, BiCurNet, is proposed for early KP estimation using spatial and harmonics domain EEG features during the biceps-curl task. The proposed model utilizes lightweight architecture with depth-wise separable convolution layers and a customized attention module. The best Pearson correlation coefficient (PCC) between the estimated and actual trajectory of $0.7$ is achieved when combined EEG features (spatial and harmonics domain) in the delta band are utilized. Intra- and Inter-subject performance analyses are performed to evaluate the subject-adaptability of the proposed decoding model. The performance of the proposed BiCurNet is compared with the existing multi-linear regression (mLR) counterpart. The robustness of the proposed model is additionally illustrated using an ablation study. The robust performance and lightweight architecture will facilitate the real-time implementation of the model with deployment on a microcontroller to control BCI-based wearable robots. 

\end{abstract}

\begin{IEEEkeywords}
Brain-computer interface, Electroencephalogram, Deep learning, Kinematic parameter estimation, Spherical harmonics, Head harmonics, Discrete wavelet transform, EEG, Feature extraction.
\end{IEEEkeywords}

\section{Introduction}
Brain-computer interface (BCI) is an integration of the measurement, decoding, and translation of the activity of the central nervous system (CNS) into imitative output that reinstates, augments, or rehabilitates the natural CNS output \cite{wolpaw2k12}. This creates an interface between the CNS and its external environment. BCI-based systems are rapidly emerging due to the recent advancements in signal processing and artificial intelligence \cite{gong2021deep,aggarwal2022review}. These systems are useful in neuro-rehabilitation to assist users with motor impairments \cite{chowdhury2017online,raza2020deep,di2020bci,li2022sensorimotor}. For real-time operability of these systems, continuous signal decoding is required to extract kinematic parameters (KPs) such as motion trajectory, velocity, acceleration, etc. \cite{jain2k22premovnet, wang2023comprehensive}. Given these aspects, electroencephalogram (EEG)-based BCI systems have gained popularity in recent years, owing to the non-invasiveness, low-cost, and excellent temporal resolution of EEG signals \cite{msainibspc2k22}.

\subsection{Motivation and Related Work}
Literary works explore machine learning and deep learning-based paradigms for upper limb kinematic parameter estimation, movement intention detection, and classification from low-frequency components of EEG signals. Sparse multinomial logistic regression is utilized to classify EEG signals during reach intention and actual movement based on multiple hand-crafted features extracted from EEG signals \cite{hammon2k7}. Independent component analysis (ICA) and dipole fitting are utilized for removing movement artifacts from the recorded EEG signals to obtain low classification error rates. Multi-linear Regression (mLR) decoder estimates the 3D trajectories of arm movement with variable velocities using EEG segments filtered in the $0.1-40$ Hz frequency range\cite{jhkim2k14}. A high correlation is reported between the movement velocities and EEG activity above the motor cortex in frontocentral and parietal areas. mLR is also utilized with $\alpha$ and $\beta$ band powers of EEG signals during the motor planning and execution phases to predict the upcoming peak tangential speed and acceleration of hand movement\cite{yang2k14}. This study demonstrates the prominence of occipital and parietal-occipital regions for the $\alpha$ band and frontal and frontal-central regions for the $\beta$ band in movement planning and execution phases. In \cite{bhagat2k16}, movement intent is decoded from movement-related cortical potentials (MRCPs) using narrow-band EEG in the range of $0.1-1$ Hz to train a support vector machine (SVM)-based classifier. Besides mLR, sparse linear regression (SLiR) is utilized for predicting the circular trajectories of the upper limb during the movement of bottles with varying masses \cite{yn2k17}. A wide range of EEG frequencies is used in this work, i.e., $0-150$ Hz and channels over the motor cortex are shown to be more prominent toward the prediction. EEG slow cortical potentials (SCPs) are utilized to decode hand, elbow, and shoulder trajectories using the mLR model during center-out-reach task\cite{sosnik2020reconstruction}. EEG current source dipole-based trajectory estimation is used for decoding actual and imagined arm joint trajectories based on multiple linear regression (mLR)\cite{rs2k21}. The most useful time lags are observed to be between $80-150$ ms before the movement, and the low $\beta$ and $\gamma$ bands are shown to be more effective in movement decoding with a correlation of $0.67$. In a recent study, researchers have explored the feasibility of a commercial EEG headset in motor decoding and classification using the Kalman filter and spatio-spectral features extracted from EEG signals \cite{nrobin2k21}. An overall correlation of $0.58$ is achieved in this work. The selection of a single-channel, i.e., $C_z$ in movement onset decoding with an accuracy of $91\%$ using low frequency ($0-5$ Hz) and Teager-Kaiser energy operator with threshold-based classification is demonstrated in \cite{mahmoodi2k21}.

Despite the effectiveness of conventional machine learning-based paradigms in EEG-based movement decoding, there is a need to extract high-level features to enhance performance. To overcome this, deep learning-based paradigms have been proposed. For example, a convolutional neural network (CNN) is proposed using pre-movement raw spatio-temporal multi-channel EEG for hand movement and force levels classification with an accuracy of $84\%$ \cite{gatti2k21}. This work demonstrates early hand movement classification, i.e., in $100-1600$ ms advance. CNN combined with a bidirectional long short-term memory (Bi-LSTM)-based network is also used to predict arm-reaching task velocities. An overall correlation between $0.4-0.6$ is achieved in this work, and the feasibility of robotic arm control based on real-time EEG is demonstrated \cite{jeong2k20}. Recently, deep learning-based three-dimensional (3D) hand movement trajectory during grasp and lift movements is estimated using a public EEG database in \cite{kumars2k21,pancholi2022source,jain2023subject}. The feasibility of a brain-inspired spiking neural network (Bi-SNN) along with mid-frequency and high-frequency EEG bands, i.e., $\alpha$, $\beta$, and $\gamma$, is demonstrated toward the trajectory estimation with a correlation of $0.7$ \cite{kumars2k21}. In \cite{pancholi2022source}, wavelet packet decomposition (WPD) based time-lagged EEG sub-bands train a CNN-LSTM network to predict the hand position/trajectory with a high correlation of $0.86$. This work explores the source-aware EEG features and demonstrates the relevance of low-frequency bands ($\delta$, $\theta$, and $\alpha$) in movement estimation. However, it has limited feasibility in real-time hardware implementation. The subject-independent trajectory estimation is demonstrated in \cite{jain2023subject} with various pre-movement EEG windows using a CNN-LSTM based decoder.

Based on the aforementioned description of literary works, it can be asserted that many of these works focus on the classification of upper limb movements rather than the prediction/estimation of the related kinematic parameters. Timely extraction of kinematic parameters from EEG data during upper limb movement is imperative for different real-time exosuit control-based BCI applications. Further, few existing machine-learning based regression algorithms can estimate the KPs earlier w.r.t. actual movement. However, an average correlation is achieved. Although the existing deep learning-based networks outperform these ML-based paradigms, only a few have explored early estimation of KPs. Further, these networks use slightly complex architectures after pre-processing, which may not be feasible on microprocessors/microcontrollers for real-time BCI systems. Most importantly, the performance of the existing paradigms for KP estimation is highly subject-specific, adding to the complexity since the networks must be trained for each subject. 

\subsection{Objective and Key Contributions}
This work proposes a deep learning-based upper limb motion trajectory prediction/estimation from preceding EEG, i.e., BiCurNet, for early estimation towards exosuit control-based BCI applications. Further, the proposed network is demonstrated to be subject-independent and robust against artifacts. The work focuses on the early estimation of kinematic parameters from subject-dependent and subject-independent EEG signals and further analyses the noise-robustness of the proposed network. The key contributions of this work are listed as follows.
\begin{itemize}
\item Data collection of multi-channel EEG signals during upper limb biceps curl experiment.
\item Spherical harmonics and head-harmonics domain EEG features based motion trajectory estimation has been explored for the first time.
\item Low-complex deep learning-based architecture is proposed for early estimation of upper limb motion trajectory.
\item Demonstration of subject-adaptability and noise-robustness of the proposed network.
\end{itemize}
The rest of this paper is organized as follows. Section II describes the experimental recording and data acquisition procedures. Section III presents the proposed methodology for BiCurNet. Section IV discusses the experimental evaluation results for the proposed work. Finally, section V concludes this work with major advantages, shortcomings, and future directions. 
\begin{figure}[t]
        \centering	\includegraphics[width=0.45\textwidth]{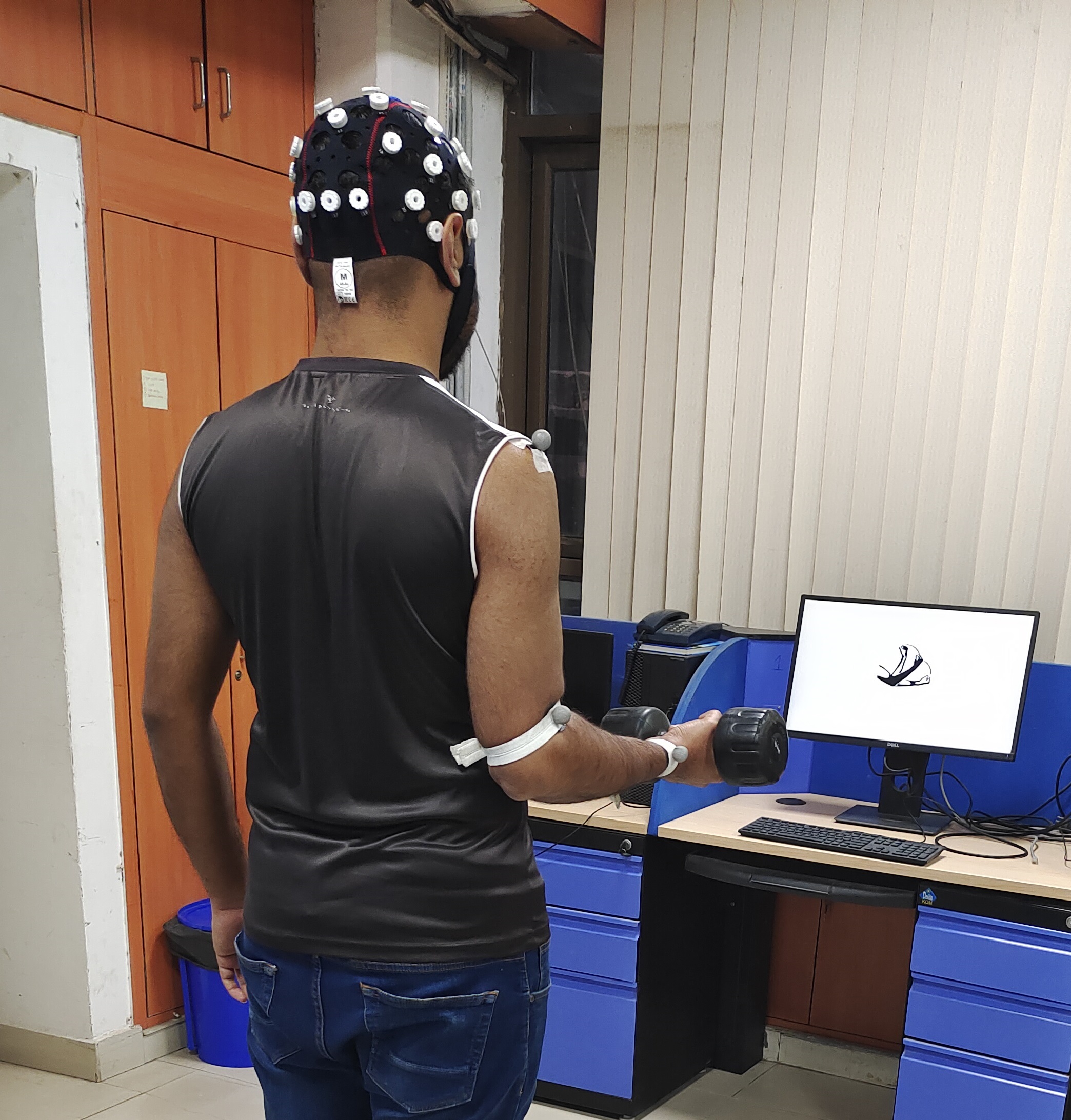}
	\caption{Experimental setup for biceps-curl task.}
	\label{exp} 
\end{figure}
\begin{figure*}[t]
\centering
\includegraphics[width=\textwidth]{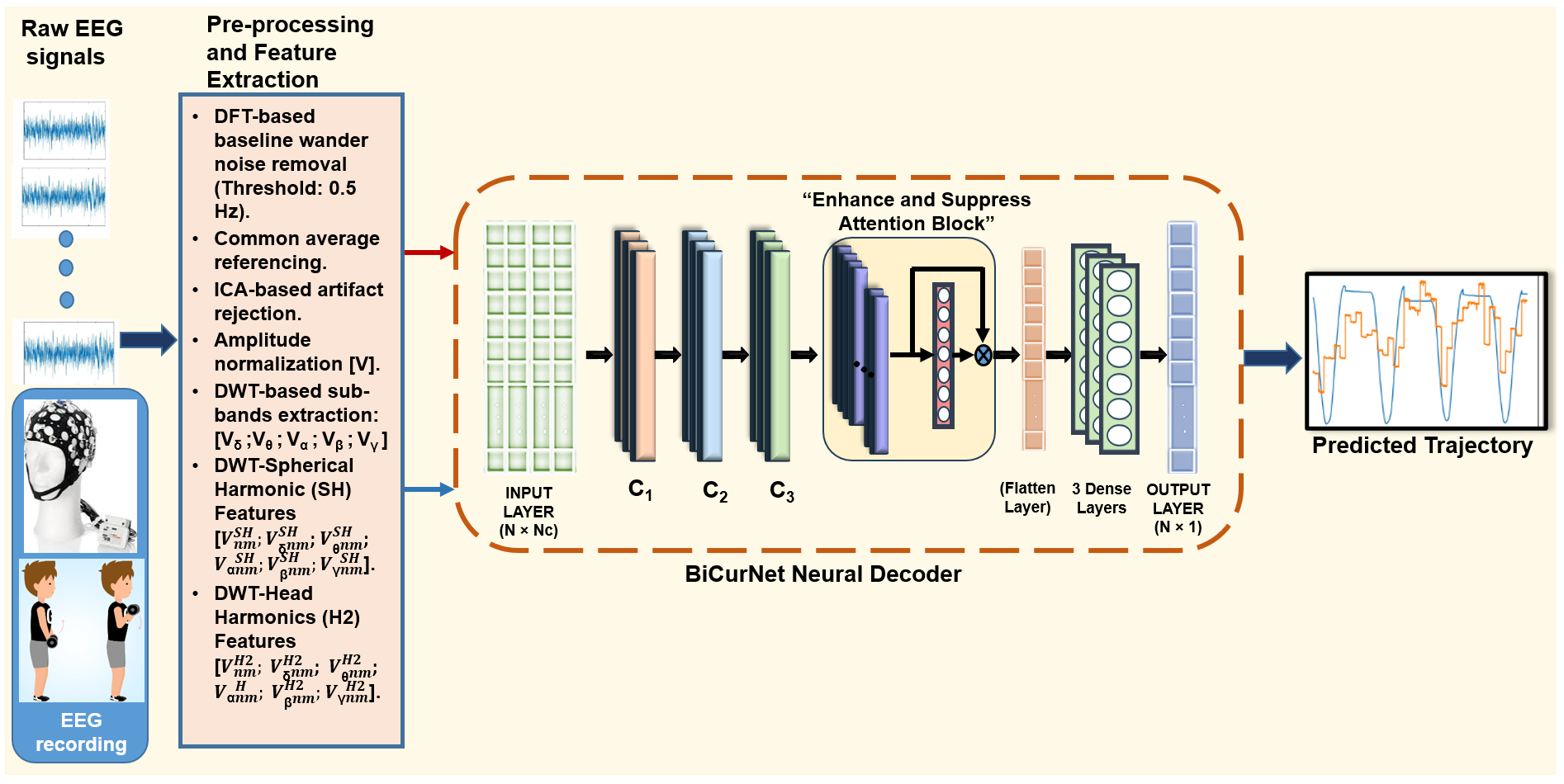}
\caption{Block diagram depicting the proposed methodology for biceps-curl trajectory estimation. The proposed neural decoding model, BiCurNet, consists of an input layer, a DWSConv1D layer ($C_1$), two Conv1D layers ($C_2$ and $C_3$), an attention module, followed by a flatten layer, three dense layers (eight neurons each) and an output layer.}
\label{fig:figmethod}
\end{figure*}
\section{Experiment and Data Acquisition}

The key objective of the study is to investigate the viability of using EEG signals for elbow joint angle decoding during biceps-curl motion. For this purpose, we designed a synchronous EEG and joint angle recording system. The description of the experimental paradigm and data acquisition are elucidated in the subsequent sections.

\subsection{Subjects and Equipment}

The experiment was performed in the Multichannel Signal Processing Laboratory, Department of Electrical Engineering at the Indian Institute of Technology Delhi, New Delhi. This research was authorized by the Institutional Review Board of All India Institute Of Medical Sciences, New Delhi. EEG and joint angle data were recorded from 5 healthy subjects (all males, age $29 \pm 2.61$, all right-handed) while performing the biceps curl task. Each subject performed 300 trials of the biceps curls task. EEG data was recorded using 16-channel dry-active electrodes (actiCAP Xpress Twist, Brain Products, Gilching, Germany) with a wireless EEG amplifier (LiveAmp-16, Brain Products, Gilching, Germany). The EEG sensors were arranged in 10-20 international systems of EEG electrode placement, namely, Fp1, Fz, F3, C3, T7, Pz, P3, O1, Oz, O2, P4, Cz, C4, T8, F4, and Fp2. The EEG data was acquired with a 500 Hz sampling frequency. A marker-based camera system (Noraxon NiNOX 125 Camera System) was placed for elbow joint angle measurement. The NiNOX 125 camera system was connected to the Noraxon myoResearch platform (MR 3.16) to record the biceps-curl hand motion. The camera system was placed in a sagittal plane 2m from the subject. The elbow joint angle was calculated using myoResearch software in the post-processing step. The 3-point angle measurement tool was utilized to compute 2D joint angle by tracking reflective markers in the video recording. The joint angle data was sampled with a sampling frequency of 125 Hz. The EEG and joint angle data were synchronized using a Noraxon myoSync device.

\subsection{Experimental Setup and Paradigm}

Concurrent EEG and motion data were collected from the users during the biceps-curl task. At the beginning of the experiment, participants were in a standing position with a 2 Kg dumbbell holding in their right hand. A monitor was positioned 1.8 m away in front of them to show the experimental paradigm as shown in Fig. \ref{exp}. Participants stood in a balanced upright posture with the dumbbell in their right hand. We designed the experiment in PsychoPy \cite{peirce2007psychopy} to instruct the participant for the initiation of biceps-curl movement. Each trial begins with a cross appearing on the center screen and a beep sound, indicating the start of the trial. After a couple of seconds, a visual cue appeared on the screen to instruct the participant to initiate the biceps curl. The biceps-curl was performed in the motion execution phase. Each trial ended with a resting phase of two seconds. Before the data acquisition, each participant performed a practice run to execute the task correctly. This practice run was not included in any consequent analysis. We recorded 30 runs with 10 trials each for the biceps curl task. Inter-run rest was given to the participant to avoid muscle fatigue.

\section{Proposed methodology}
This section elaborates on the proposed methodology for early prediction of upper limb motion trajectory from EEG based on deep learning, as illustrated in Fig. \ref{fig:figmethod}. It consists of three major modules: EEG recording, pre-processing and feature extraction, and depth-wise separable convolutional neural network with a customized attention module. The modules are described in the subsequent sub-sections.

\subsection{EEG recording}
In this work, the EEG signals are acquired using the LiveAmp-16 Brain Products system, as described in the previous section. EEG signals are pre-processed before input for the proposed BiCurNet, as detailed in the ensuing sub-section.

\begin{figure*}[t]
	\centering
	\subfigure[]{\includegraphics[width=0.23\textwidth,height=0.20\textwidth]{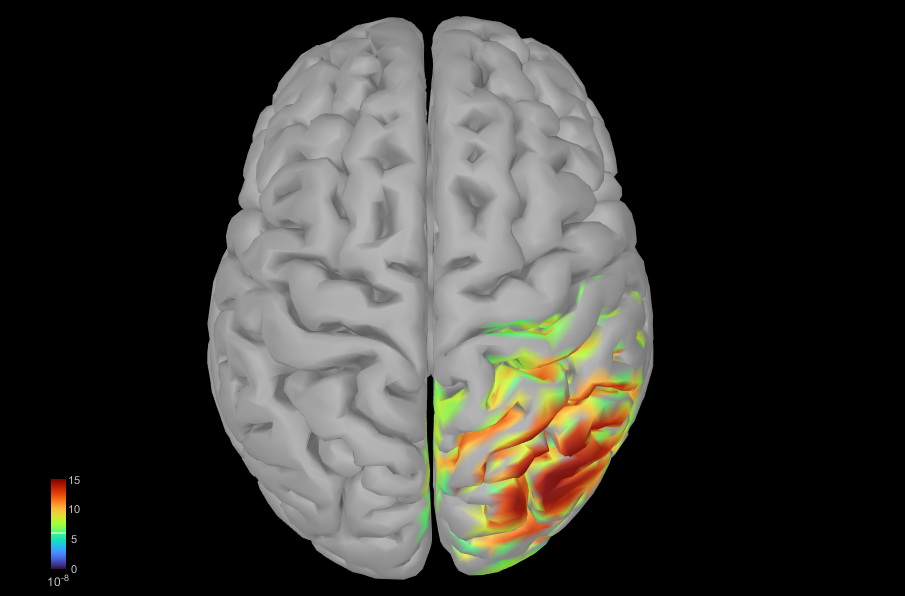}}
	\subfigure[]{\includegraphics[width=0.23\textwidth,height=0.20\textwidth]{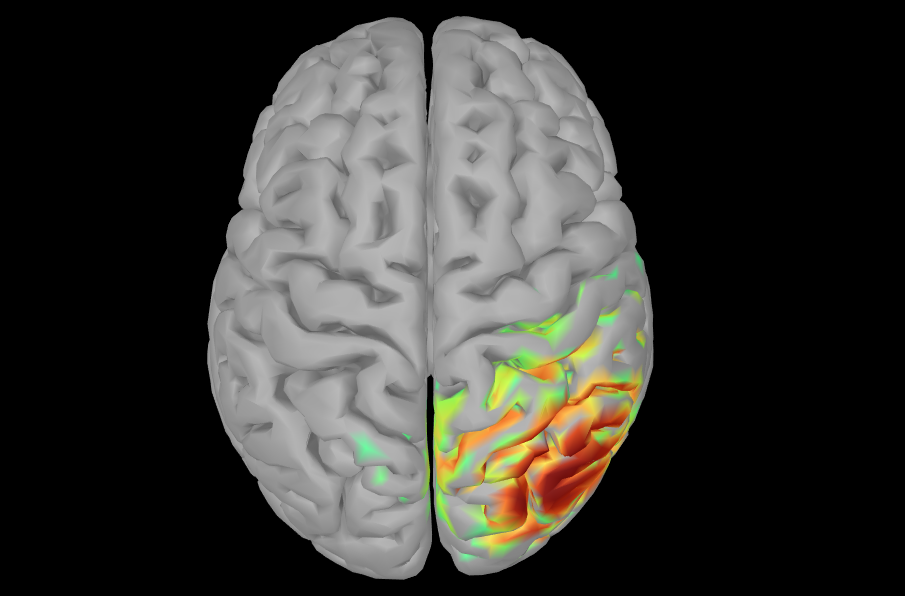}}
	\subfigure[]{\includegraphics[width=0.23\textwidth,height=0.20\textwidth]{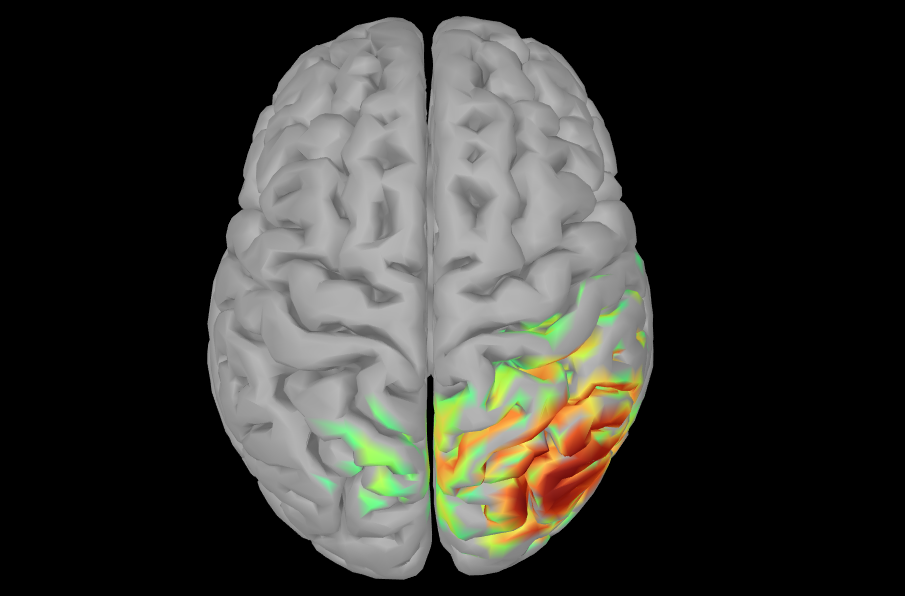}}
	\subfigure[]{\includegraphics[width=0.23\textwidth,height=0.20\textwidth]{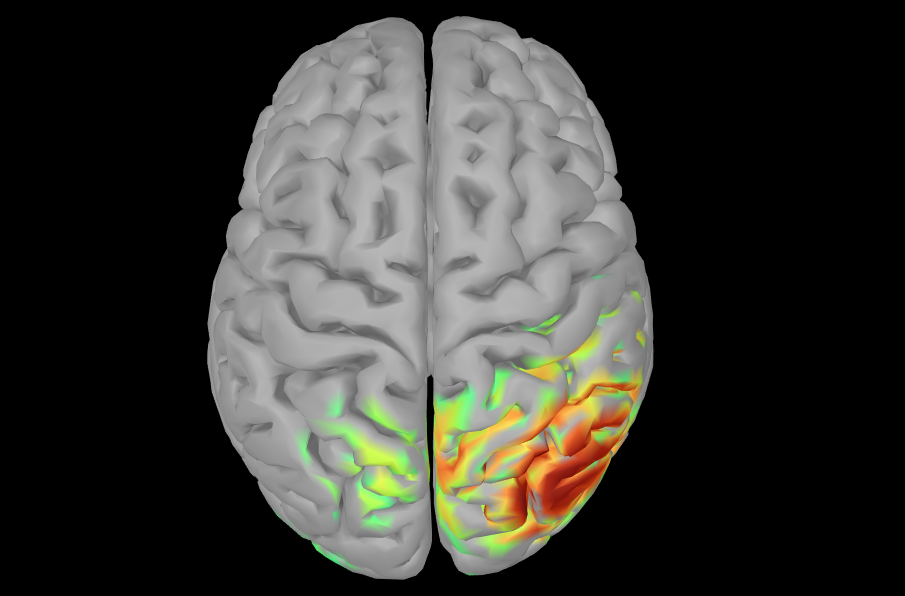}}
	\subfigure[]{\includegraphics[width=0.23\textwidth,height=0.20\textwidth]{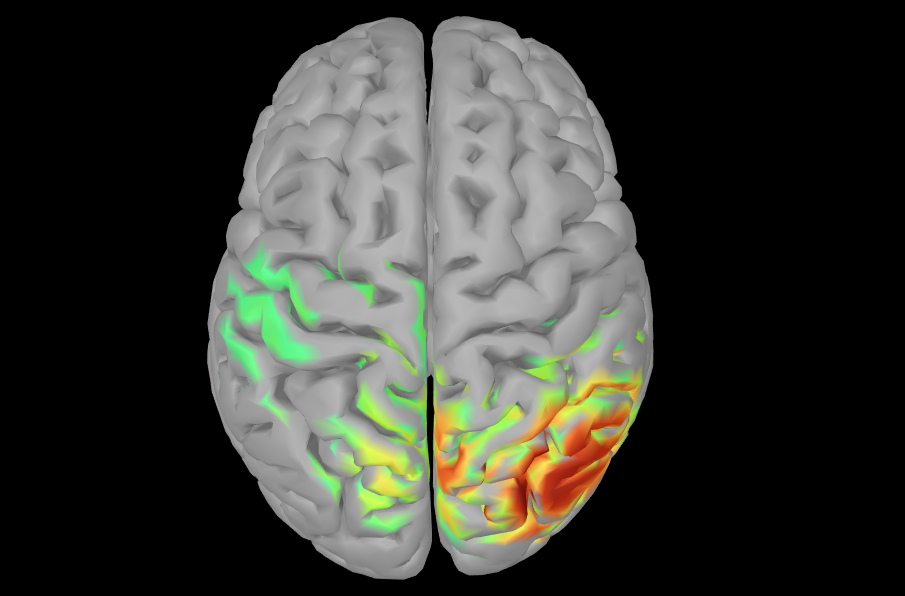}}
	\subfigure[]{\includegraphics[width=0.23\textwidth,height=0.20\textwidth]{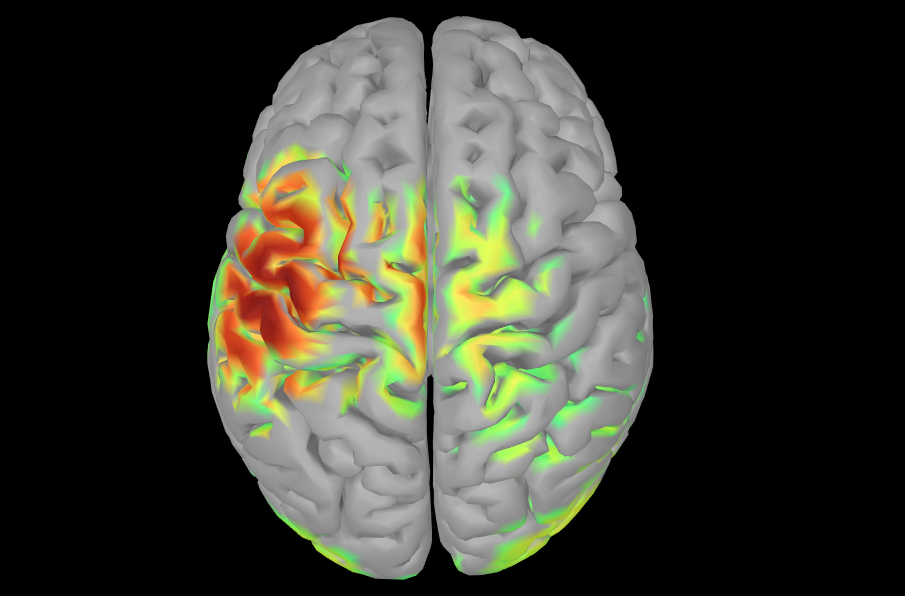}}
	\subfigure[]{\includegraphics[width=0.23\textwidth,height=0.20\textwidth]{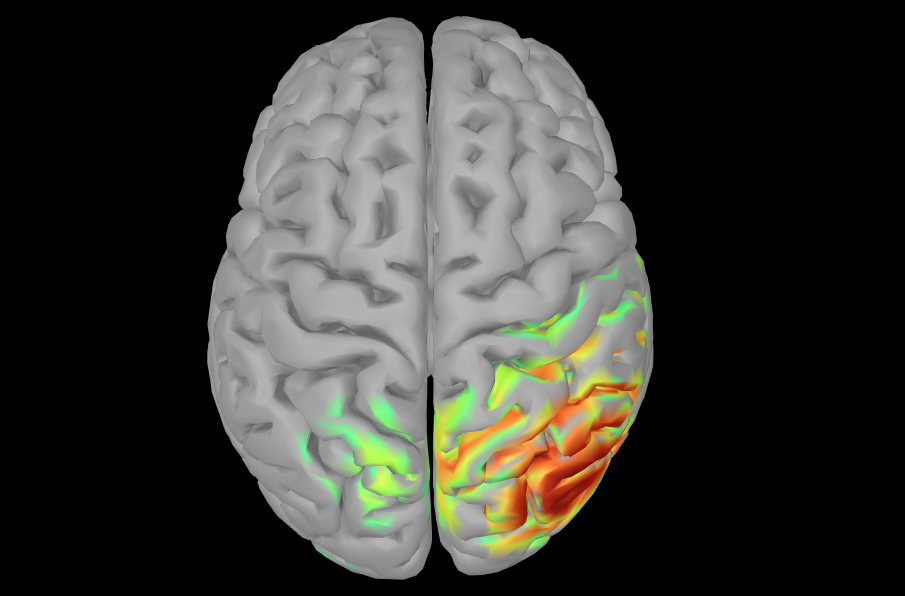}}
	\subfigure[]{\includegraphics[width=0.23\textwidth,height=0.20\textwidth]{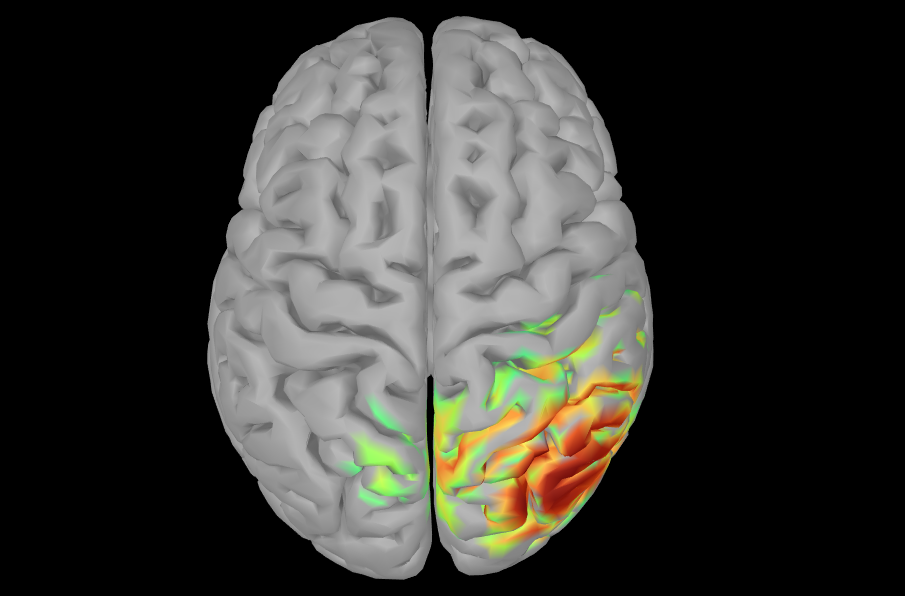}}
	\caption{Brain source localization using sLORETA at different time stamps : (a) 0ms (b) 60ms (c) 120ms (d) 180ms (e) 240ms (f) 300ms (g) 360ms (h) 420ms} 
	\label{sloreta_maps}
	
\end{figure*}

\subsection{Pre-processing}
The recorded EEG signals are pre-processed in EEGLAB \cite{delorme2004eeglab} for feature extraction before being fed to the proposed BiCurNet, as shown in Fig. \ref{fig:figmethod}. After recording and re-sampling the EEG signals, low frequency (below $0.5$ Hz) baseline wander noise (BWN) suppression is done using discrete Fourier transform (DFT). For this purpose, the DFT coefficients corresponding to frequencies below $0.5 Hz$ are estimated. The computation of DFT coefficient index $k$ is done as $k=\lfloor(f_q N_d / f_{qs})\rfloor$, where $f_{q}$ is the frequency in $Hz$, $f_{qs}$ is the sampling frequency, and $N_d$ is the number of DFT points for computation. These DFT coefficients are thresholded to zero for suppression of the BWN. The EEG signal after BWN suppression is synthesized as the inverse of the DFT coefficient matrix. The mathematical interpretation of this procedure is described for a recorded EEG signal $v[n]$ by the following DFT pair:
\begin{equation}
\text{DFT of recorded signal}: V[k]=\sum_{n=0}^{N_d-1} v[n] e^{\frac{-j n 2\pi k}{N_d}}
\end{equation}
\begin{equation}
\tilde{v_q}[n]=\frac{1}{N_d} \sum_{k=0}^{N_d-1} \tilde{V_q}(k) e^{\frac{j n 2\pi k}{N_d}}
\end{equation}
where $\tilde{X_q}$ denotes the DFT coefficient matrix after thresholding, i.e., $\tilde{X_q}(k)=[0, \ldots, 0, X_q[k+1], \ldots, X_q[N_d-k-1], 0, . . . ., 0]$. All signals are normalized w.r.t. amplitude to bring it in range: [$-1$, $1$] as $\frac{\tilde{x_q}[n]}{\text{max}|\tilde{x_q}[n]|}$.

In this work, the recorded EEG signals are analyzed to estimate motion trajectory with and without artifact suppression. Independent component analysis (ICA) is utilized for artifact suppression. It estimates the sources corresponding to cerebral and non-cerebral activities resulting in the scalp EEG \cite{veluvolu2k22}. EEGLAB is used for the ICA-based decomposition of the EEG signals obtained after BWN removal. The decomposed independent sources with more than $70\%$ of artifactual components are rejected, and the artifact-free EEG signal is reconstructed.

\subsection{Brain source imaging}
Brain source imaging (BSI) is performed to select the relevant pre-movement EEG segment prior to feature extraction. Numerical Boundary Element Method (BEM) based forward modeling is utilized for this purpose. The head model utilizes ICBM MRI template \cite{icbm152} in OpenMEEG \cite{openmeeg} toolbox. The spatio-temporal dynamics of brain cortical sources are obtained using inverse modeling. In particular, standardized low-resolution electromagnetic tomography (sLORETA) \cite{pascual2002standardized} is utilized to solve the under-determined inverse problem. Under the constraint of the smooth source distribution, standardized current density maps are utilized for localization inference.

Source localization plots for a right-hand biceps-curl activity are illustrated in Fig. \ref{sloreta_maps}. The analysis shown corresponds to a single trial of biceps-curl. The subject was instructed to focus the vision on the fixation cross. A visual cue for movement onset was presented at 0 ms. The subject executed biceps-curl activity 410 ms after the visual cue was given. A constant activation may be observed in the occipital lobe up to 240 ms. The information starts getting transferred to the left motor cortex thereafter. All such pre-movement EEG [Fig. \ref{sloreta_maps}(c)-(g)] has inherent information of motor execution.

It may be noted that the left motor cortex region activation was initiated at 220-240 ms [Fig. \ref{sloreta_maps}(e)] corresponding to the right-hand biceps-curl activity. Motor activation was observed thereafter up to 320 msec [Fig. \ref{sloreta_maps}(f)]. After the visual cue was given, the subject executed biceps-curl activity at 400-450 ms. It may be concluded that the motor neural information corresponding to the biceps-curl activity is present approximately 250 ms prior to the motor execution. This information is utilized for selecting the time-lag window for elbow joint-angle trajectory. The selected EEG data was utilized to train and test the proposed neural decoder.

\subsection{Feature extraction}
The pre-processed EEG signals are analyzed with different transform-domain techniques for significant feature extraction. This work explores time-frequency features using discrete wavelet transform (DWT) in the Spatial domain, and the spatio-temporal domain using spherical Fourier transform (SFT) and head harmonics transform.

\subsubsection{Discrete wavelet transform-based features}
Discrete wavelet transform decomposes the EEG signals into constituent sub-bands/rhythms. It uses high-pass and low-pass filters to decompose the signals into a pre-defined number of levels based on the sampling frequency \cite{msainispringer2k22}. DWT of a single channel EEG signal $v[n]$ is given by
\begin{equation}
V_{j, r}=\sum_{n \in z} v[n] \psi_{j, r}^{*}[n]
\label{Eq11}
\end{equation}
where ${\psi_{j, r}}$ is the translated and scaled version of the mother wavelet $\psi_{0,0}$, and defined as:
\begin{equation}
\psi_{j, r}[n]=2^{-(j / 2)} \psi_{0,0}\left(2^{-j}(n-r)\right)
\end{equation}

\begin{figure}[t]
\includegraphics[width=\linewidth]{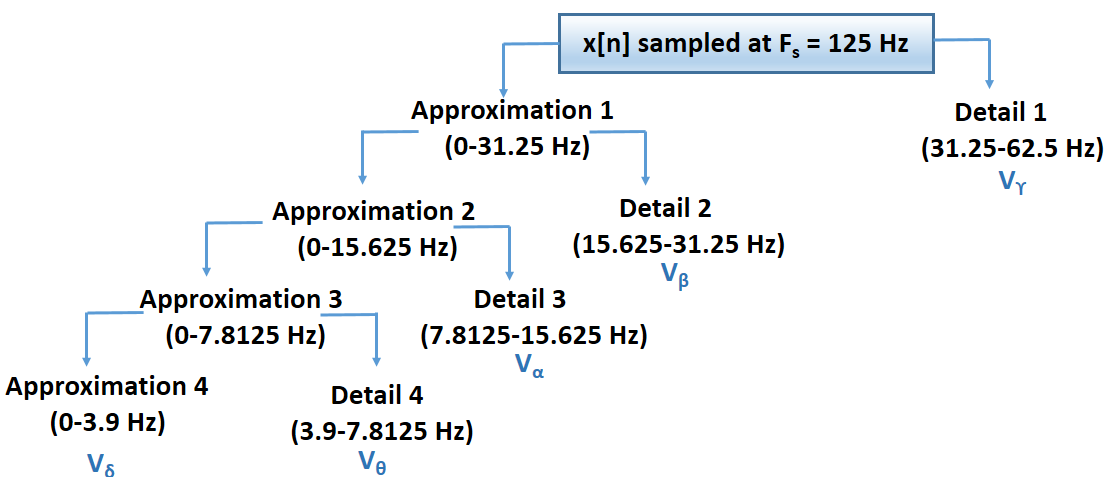}
\caption{Four-level DWT-based decomposition to obtain the approximation and detail bands with frequency range at level $j$ given by:
 $\left[0,2^{-j-1} \mathrm{~F}_s\right] \text {and }\left[2^{-j-1} \mathrm{~F}_s, 2^{-j} \mathrm{~F}_s\right]$, respectively.}
\label{fig:figdwt}
\end{figure}

The procedure for DWT-based decomposition follows a tree-like structure as demonstrated in Fig. \ref{fig:figdwt}. The wavelet coefficients are down-sampled at each decomposition level to remove the redundant information. {\color{red}A four-level DWT-based decomposition is performed such that the last approximation level result in the delta band frequency range\cite{mallat}. In this work, since the sampling frequency used is $125$ Hz, the decomposed sub-bands are obtained as delta ($\delta: 0.5-3.9$ Hz), theta ($\theta: 3.9-7.8$ Hz), alpha ($\alpha: 7.8-15.6$ Hz), beta ($\beta: 15.6-31.2$ Hz), and gamma ($\gamma: >31.2$ Hz), denoted by $V_{\delta}$, $V_{\theta}$, $V_{\alpha}$, $V_{\beta}$, and $V_{\gamma}$ respectively. }

\subsubsection{DWT-Spherical harmonics-based features} \label{dwt-sh}
\begin{figure}[h]
\centering
\includegraphics[width=0.65\linewidth]{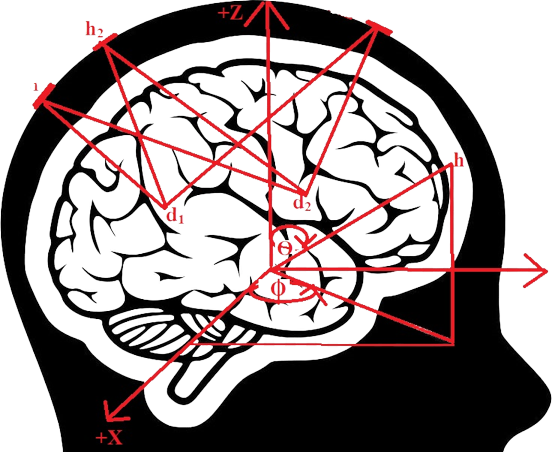}
\caption{Total potential at a channel contributes to each active equivalent dipole.}
\label{fig:figsfd}
\end{figure}

To extract the spatio-temporal features of the EEG signal and the corresponding DWT-based sub-bands obtained above, the spherical Fourier transform (SFT) is explored in this work. Since the human head is assumed to be spherical in shape \cite{agiri2k20}, spherical Fourier basis functions have been widely employed in literary works. The decomposition of a multi-channel EEG signal $\mathbf V$ in SFD is obtained as:
\begin{equation}
\mathbf V_{l m}^{SH} = \int_{\Omega} \mathbf V\left(\Omega, n\right)\left[Y_n^m(\Omega)\right] d \Omega
\end{equation}
where $V\left(\Omega, n\right)$ denotes the potential at ($\Omega$) = ($r,\theta,\phi$) on the scalp at time instant n. Here, $r$ represents the radius of the head, $\theta$ denotes the angle of elevation measured in the downward direction from the positive Z-axis ($\theta \in[0, \pi]$), and $\phi$ denotes the azimuth angle measured in anticlockwise direction from positive X-axis, as shown in Fig. \ref{fig:figsfd}. The real-valued $Y_l^m(\Omega)$ of $l^{th}$ order and $m^{th}$ degree constitute an orthonormal set of basis functions defined over the spherical array. 
For a finite order system, $l \in [0, L]$, and $m \in [-l, l]$. Therefore, $(L + 1)^2$ distinct spherical harmonics are obtained. Since the number
of sampling points $S$ in the spatial domain should be at least $(L + 1)^2$, the highest limit of L is $ \leq \sqrt(S)- 1$. In this work, since $16$ electrodes are used for recording, i.e., $S = 16$, the limit of $L$ is $3$. Therefore, $L=2$ is used here, and $9$ distinct spherical harmonics are obtained. The corresponding features are stored in $V^{SH}_{nm}$ with a dimension of $9 \times N$. Each EEG sub-band is also decomposed using spherical Fourier basis functions, and the corresponding features are obtained as $V_{\delta_{lm}}^{SH}$, $V_{\theta_{lm}}^{SH}$, $V_{\alpha_{lm}}^{SH}$, $V_{\beta_{lm}}^{SH}$, and $V_{\gamma_{lm}}^{SH}$.

\subsubsection{DWT-Head harmonics-based features}
\begin{figure}[h]
\includegraphics[width=\linewidth]{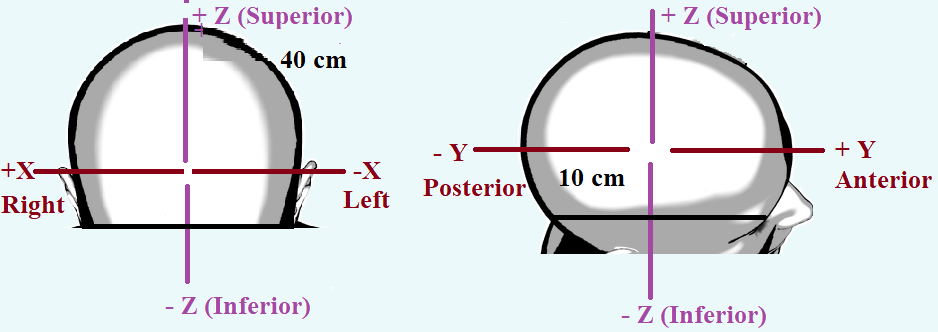}
\caption{Geometry of human head with the parameters: Perimeter=40cm, radius=10cm.}
\label{fig:fighhd}
\end{figure}

More recently, head harmonics ($H^2$) basis functions have been proposed for an adequate representation of EEG signals based on the geometry of the human head \cite{agiri2k20}. Since the EEG sensors placed on the head form a shape between a sphere and a hemisphere, $H^2$ basis functions are shown to be more efficient for representing the data sampled overhead. The decomposition of an EEG signal matrix $\mathbf V$ in $H^2$ domain is given as:
\begin{equation}
\begin{aligned}
\mathbf V_{l m}^{H2} &=\int_{\Omega} \mathbf V\left(\Omega, n\right)\left[H_l^m(\Omega)\right] d \Omega \\
& \approx \sum_{w=1}^S z_w V\left(\Omega_w, n\right)\left[H_l^m\left(\Omega_w\right)\right]
\end{aligned}
\end{equation}
where, $z_w$ denotes the sampling weight and $\Omega_w=\left(\theta_w, \phi_w\right)$ is the location of channel $w$. Here, the elevation angle $\theta$ is in the range $[0, 2\pi/3]$, as per the head geometry shown in Fig. \ref{fig:fighhd}. The real-valued $H_l^m(\Omega)$ of $l^{th}$ order and $m^{th}$ degree constitute an orthonormal set of basis functions defined over the human head. 

The corresponding features are stored in $V^{H2}_{lm}$ with a dimension of $9 \times N$, similar to that obtained in SFT. Each EEG sub-band is also decomposed using $H^2$ basis functions, and the corresponding features are obtained as $V_{\delta_{lm}}^{H2}$, $V_{\theta_{lm}}^{H2}$, $V_{\alpha_{lm}}^{H2}$, $V_{\beta_{lm}}^{H2}$, and $V_{\gamma_{lm}}^{H2}$.

\subsection{Proposed BiCurNet}
After pre-processing and feature extraction, the EEG data is given as input to the proposed BiCurNet model. The proposed deep learning model is illustrated in Fig. \ref{fig:figmethod}. The constituent layers in the proposed model include a depth-wise separable one-dimensional convolution layer (DWSConv1D), a conv1D layer, a max-pooling (maxpool1D) layer, a customized attention module (CAM), a flatten layer, three dense layers, and an output layer.
\begin{itemize}
\item Depth-wise separable convolution layer (DWSConv1D): The first layer of the network is a conv1D layer which performs a depth-wise separable convolution of the received input data with the kernels/filters used in this layer. It receives the input EEG data in the form of $N \times Nc$ matrix as shown in Fig. \ref{fig:figmethod}. Here $N$ denotes the number of samples in the data, and $Nc$ is the number of channels. The convolution operation is split into two parts in this layer, depth-wise and point-wise \cite{cbam2k18}. Depth-wise convolution is performed with each channel separately, and point-wise convolution is performed as $1\times1$ convolution. {\color{red}It is a computationally efficient operation w.r.t. the standard convolution layer, making it suitable for lightweight scenarios.} Convolution of a filter $f[n]$ with an input $v[n]$ is written as:
\begin{equation}
v[n] \ast f[n]=\sum_{i=0}^{k_s-1} v[i] \cdot f[n-i]
\end{equation}
where, `$\ast$' represents the convolution operation and $k_s$ denotes the filter width. In this layer, $32$ filters are used. Each filter has a width $k_s$ of $5$.
In general, the $z^{th}$ convolution output, i.e, feature map of layer $lr$ is given as \cite{prez}:
\begin{equation}
c_{z}^{lr}=\sigma\left(bi_{z}^{lr}+\sum_{j} c_{j}^{lr-1} \times f_{z j}^{lr}\right)
\end{equation}
where $c_{z}^{lr}$ is the $z^{th}$ feature in the $lr^{th}$ layer; $c_{j}^{lr-1}$ is the $j^{th}$ feature in the corresponding preceding layer; $f_{z j}^{lr}$ represents the filter which links feature $z$ to feature $j$, $bi_{z}^{lr}$ represents the corresponding bias vector and $\sigma$ denotes the activation function, which is rectified linear unit (ReLu) in this layer. 
It is defined as: $\sigma(t)=\max (0, t)$. A stride of one is used in this layer. The 'He' uniform initialization is used for kernel weights, and zero initialization is used for the bias vector. All these parameters produce an output dimension of $C_1$: $(N-k_s+1) \times 32$ as shown in Fig. \ref{fig:figmethod}. $L2$ regularization with a factor of $0.001$ is also used in this layer to reduce over-fitting.
\item Conv1D layer: The second layer is a conventional convolution layer that simultaneously operates on all input channels. This layer uses the same parameters as described in the previous layer. The corresponding output dimension of this layer is given as $(C_1-k_s+1) \times 32$.


\item Max pooling layer (Maxpool1D): The convolution layer output is reduced in dimensionality by using a max pooling 1D layer, which retains the highest value of the feature in a segment with a pool size \cite{jiao}. This layer helps in low-level feature extraction. The corresponding process can be interpreted as \cite{prez}:
\begin{equation}
c_{mx}^{hx}=\max _{\forall b \in ar_{m}} c_{b}^{hx-1}
\end{equation}
where, $ar_{m}$ denotes the pool area with index $m$. 
In this work, a pool size and a stride of $2$ are selected, which results in the dimension of the output as $(C_1-k_s+1)/2 \times 32$, shown in Fig. \ref{fig:figmethod}.

\item Customized attention module (CAM): {\color{red}The feature maps of the previous layer are further transformed to intensify the more relevant features and restrain the less relevant features.} A CAM is utilized for this purpose, which uses a dense layer with $32$ units and a multiply layer, as shown in Fig. \ref{fig:figmethod}. This module works on the attention phenomenon, which enhances the relevant features and diminishes the less significant features \cite{cbam2k18}. An element-wise multiplication operation is performed between the outputs of the dense layer and the maxpool1D layer. This produces higher values of product where both maxpool1D and dense layer outputs are high, thereby enhancing the more intense features. Similarly, the less significant features are further restrained due to low product values where both the layer outputs are low. The input dimension of the dense layer is $(C_3) \times 32$, and a dot product operation between a $32 \times 32$ weight vector of the dense layer and its input results in the same output dimension.
\item Flatten layer: This layer transforms the output of CAM, $C_3 \times 32$, to a 1D vector with dimension $C_4 \times 1$, as shown in Fig. \ref{fig:figmethod}. A dropout with a factor of $0.4$ is used after this layer to prevent the model from over-fitting \cite{wu}.
%
\begin{table}
\vspace{-0.45cm}
\caption{Training hyper-parameters (After hyper tuning).}
\large
\scalebox{0.8}{
\begin{tabular}{|m{0.8cm}|m{0.8cm}|m{0.8cm}|m{0.8cm}|m{0.8cm}|m{1.2cm}|m{1cm}|m{0.8cm}|m{0.8cm}|}

\hline
         $N_{c}$  & $N_{k}$  & $D_r$ & $k_s$  &
        $s_r$  &  $l_r$  & $B_t$  & $e_c$ \\ \hline
         3 & 32 & 0.40  & 5 &1 & 0.001 & 15 & 100 \\ \hline
         
\end{tabular}}
\label{Tabhyper}
\vspace{0.1cm}\\
\scriptsize{ $N_{c}$: Number of convolution layers, $N_{k}$: Number of kernels/filters, $D_r$: Dropout rate, $k_{s}$: Kernel width, $s_r$: Stride/shift, $l_r$: Learning rate, $B_t$: Batch size, $e_c$: Number of training epochs. }
\end{table}
\begin{table*}[ht]
\centering
\caption{{Pearson correlation coefficient (PCC) for different EEG segments and lags of data (Mean over subjects).}}
\newcommand{\cc}[1]{\multicolumn{1}{c}{#1}}
\renewcommand{\tabcolsep}{0.5pc} 
\renewcommand{\arraystretch}{1.2}
\large
\scalebox{0.69}{
\begin{tabular}{|m{1.5cm}||m{0.6cm}|m{0.7cm}|m{0.7cm}|m{0.85cm}|m{0.85cm}||m{0.6cm}|m{0.7cm}|m{0.7cm}
                |m{0.85cm}|m{0.85cm}||m{0.6cm}|m{0.7cm}|m{0.7cm}|m{0.85cm}|m{0.85cm}||m{0.6cm}|m{0.7cm}
                |m{0.7cm}|m{0.85cm}|m{0.85cm}|}
\hline
\rowcolor[rgb]{ .851,  .851,  .851} EEG Features & {\cellcolor[rgb]{ .886,  .937,  .855}8 ms} & {\cellcolor[rgb]{ .886,  .937,  .855}40 ms} & {\cellcolor[rgb]{ .886,  .937,  .855}80 ms} & {\cellcolor[rgb]{ .886,  .937,  .855}160 ms} & {\cellcolor[rgb]{ .886,  .937,  .855}240 ms} & {\cellcolor[rgb]{ .867,  .922,  .969}8 ms} & {\cellcolor[rgb]{ .867,  .922,  .969}40 ms} & {\cellcolor[rgb]{ .867,  .922,  .969}80 ms} & {\cellcolor[rgb]{ .867,  .922,  .969}160 ms} & {\cellcolor[rgb]{ .867,  .922,  .969}240 ms} & {\cellcolor[rgb]{ 1,  .949,  .8}8 ms} & {\cellcolor[rgb]{ 1,  .949,  .8}40 ms} & {\cellcolor[rgb]{ 1,  .949,  .8}80 ms} & {\cellcolor[rgb]{ 1,  .949,  .8}160 ms} & {\cellcolor[rgb]{ 1,  .949,  .8}240 ms} & {\cellcolor[rgb]{ .973,  .816,  .8}8 ms} & {\cellcolor[rgb]{ .973,  .816,  .8}40 ms} & {\cellcolor[rgb]{ .973,  .816,  .8}80 ms} & {\cellcolor[rgb]{ .973,  .816,  .8}160 ms} & {\cellcolor[rgb]{ .973,  .816,  .8}240 ms} \\
\hline \hline
\rowcolor[rgb]{ .851,  .851,  .851} $V$ & \cellcolor[rgb]{ .886,  .937,  .855}0.25 & \cellcolor[rgb]{ .886,  .937,  .855}0.25 & \cellcolor[rgb]{ .886,  .937,  .855}0.26 & \cellcolor[rgb]{ .886,  .937,  .855}0.26 & \cellcolor[rgb]{ .886,  .937,  .855}0.26 & \cellcolor[rgb]{ .867,  .922,  .969}0.35 & \cellcolor[rgb]{ .867,  .922,  .969}0.35 & \cellcolor[rgb]{ .867,  .922,  .969}0.36 & \cellcolor[rgb]{ .867,  .922,  .969}0.35 & \cellcolor[rgb]{ .867,  .922,  .969}0.26 & \cellcolor[rgb]{ 1,  .949,  .8}0.42 & \cellcolor[rgb]{ 1,  .949,  .8}0.42 & \cellcolor[rgb]{ 1,  .949,  .8}0.42 & \cellcolor[rgb]{ 1,  .949,  .8}0.42 & \cellcolor[rgb]{ 1,  .949,  .8}0.43 & \cellcolor[rgb]{ .973,  .816,  .8}0.55 & \cellcolor[rgb]{ .973,  .816,  .8}0.55 & \cellcolor[rgb]{ .973,  .816,  .8}0.55 & \cellcolor[rgb]{ .973,  .816,  .8}0.55 & \cellcolor[rgb]{ .973,  .816,  .8}0.56 \\
\hline
\rowcolor[rgb]{ .851,  .851,  .851} $V_{\delta}$ & \cellcolor[rgb]{ .886,  .937,  .855}0.34 & \cellcolor[rgb]{ .886,  .937,  .855}0.33 & \cellcolor[rgb]{ .886,  .937,  .855}0.33 & \cellcolor[rgb]{ .886,  .937,  .855}0.34 & \cellcolor[rgb]{ .886,  .937,  .855}0.36 & \cellcolor[rgb]{ .867,  .922,  .969}0.41 & \cellcolor[rgb]{ .867,  .922,  .969}0.41 & \cellcolor[rgb]{ .867,  .922,  .969}0.42 & \cellcolor[rgb]{ .867,  .922,  .969}0.42 & \cellcolor[rgb]{ .867,  .922,  .969}0.42 & \cellcolor[rgb]{ 1,  .949,  .8}0.48 & \cellcolor[rgb]{ 1,  .949,  .8}0.48 & \cellcolor[rgb]{ 1,  .949,  .8}0.48 & \cellcolor[rgb]{ 1,  .949,  .8}0.48 & \cellcolor[rgb]{ 1,  .949,  .8}0.49 & \cellcolor[rgb]{ .973,  .816,  .8}0.61 & \cellcolor[rgb]{ .973,  .816,  .8}0.61 & \cellcolor[rgb]{ .973,  .816,  .8}0.61 & \cellcolor[rgb]{ .973,  .816,  .8}0.66 & \cellcolor[rgb]{ .973,  .816,  .8}0.67 \\
\hline
\rowcolor[rgb]{ .851,  .851,  .851} $V_{\theta}$ & \cellcolor[rgb]{ .886,  .937,  .855}0.24 & \cellcolor[rgb]{ .886,  .937,  .855}0.23 & \cellcolor[rgb]{ .886,  .937,  .855}0.23 & \cellcolor[rgb]{ .886,  .937,  .855}0.24 & \cellcolor[rgb]{ .886,  .937,  .855}0.26 & \cellcolor[rgb]{ .867,  .922,  .969}0.38 & \cellcolor[rgb]{ .867,  .922,  .969}0.38 & \cellcolor[rgb]{ .867,  .922,  .969}0.38 & \cellcolor[rgb]{ .867,  .922,  .969}0.38 & \cellcolor[rgb]{ .867,  .922,  .969}0.38 & \cellcolor[rgb]{ 1,  .949,  .8}0.44 & \cellcolor[rgb]{ 1,  .949,  .8}0.44 & \cellcolor[rgb]{ 1,  .949,  .8}0.44 & \cellcolor[rgb]{ 1,  .949,  .8}0.44 & \cellcolor[rgb]{ 1,  .949,  .8}0.45 & \cellcolor[rgb]{ .973,  .816,  .8}0.55 & \cellcolor[rgb]{ .973,  .816,  .8}0.55 & \cellcolor[rgb]{ .973,  .816,  .8}0.55 & \cellcolor[rgb]{ .973,  .816,  .8}0.56 & \cellcolor[rgb]{ .973,  .816,  .8}0.57 \\
\hline
\rowcolor[rgb]{ .851,  .851,  .851} $V_{\alpha}$ & \cellcolor[rgb]{ .886,  .937,  .855}0.22 & \cellcolor[rgb]{ .886,  .937,  .855}0.22 & \cellcolor[rgb]{ .886,  .937,  .855}0.22 & \cellcolor[rgb]{ .886,  .937,  .855}0.22 & \cellcolor[rgb]{ .886,  .937,  .855}0.21 & \cellcolor[rgb]{ .867,  .922,  .969}0.36 & \cellcolor[rgb]{ .867,  .922,  .969}0.36 & \cellcolor[rgb]{ .867,  .922,  .969}0.37 & \cellcolor[rgb]{ .867,  .922,  .969}0.36 & \cellcolor[rgb]{ .867,  .922,  .969}0.36 & \cellcolor[rgb]{ 1,  .949,  .8}0.39 & \cellcolor[rgb]{ 1,  .949,  .8}0.39 & \cellcolor[rgb]{ 1,  .949,  .8}0.39 & \cellcolor[rgb]{ 1,  .949,  .8}0.39 & \cellcolor[rgb]{ 1,  .949,  .8}0.39 & \cellcolor[rgb]{ .973,  .816,  .8}0.51 & \cellcolor[rgb]{ .973,  .816,  .8}0.51 & \cellcolor[rgb]{ .973,  .816,  .8}0.51 & \cellcolor[rgb]{ .973,  .816,  .8}0.51 & \cellcolor[rgb]{ .973,  .816,  .8}0.53 \\
\hline
\rowcolor[rgb]{ .851,  .851,  .851} $V_{\beta}$ & \cellcolor[rgb]{ .886,  .937,  .855}0.18 & \cellcolor[rgb]{ .886,  .937,  .855}0.18 & \cellcolor[rgb]{ .886,  .937,  .855}0.17 & \cellcolor[rgb]{ .886,  .937,  .855}0.17 & \cellcolor[rgb]{ .886,  .937,  .855}0.17 & \cellcolor[rgb]{ .867,  .922,  .969}0.29 & \cellcolor[rgb]{ .867,  .922,  .969}0.29 & \cellcolor[rgb]{ .867,  .922,  .969}0.29 & \cellcolor[rgb]{ .867,  .922,  .969}0.3 & \cellcolor[rgb]{ .867,  .922,  .969}0.3 & \cellcolor[rgb]{ 1,  .949,  .8}0.32 & \cellcolor[rgb]{ 1,  .949,  .8}0.32 & \cellcolor[rgb]{ 1,  .949,  .8}0.32 & \cellcolor[rgb]{ 1,  .949,  .8}0.32 & \cellcolor[rgb]{ 1,  .949,  .8}0.33 & \cellcolor[rgb]{ .973,  .816,  .8}0.39 & \cellcolor[rgb]{ .973,  .816,  .8}0.39 & \cellcolor[rgb]{ .973,  .816,  .8}0.39 & \cellcolor[rgb]{ .973,  .816,  .8}0.39 & \cellcolor[rgb]{ .973,  .816,  .8}0.39 \\
\hline
\rowcolor[rgb]{ .851,  .851,  .851} $V_{\gamma}$ & \cellcolor[rgb]{ .886,  .937,  .855}0.1 & \cellcolor[rgb]{ .886,  .937,  .855}0.1 & \cellcolor[rgb]{ .886,  .937,  .855}0.1 & \cellcolor[rgb]{ .886,  .937,  .855}0.1 & \cellcolor[rgb]{ .886,  .937,  .855}0.1 & \cellcolor[rgb]{ .867,  .922,  .969}0.17 & \cellcolor[rgb]{ .867,  .922,  .969}0.17 & \cellcolor[rgb]{ .867,  .922,  .969}0.17 & \cellcolor[rgb]{ .867,  .922,  .969}0.18 & \cellcolor[rgb]{ .867,  .922,  .969}0.18 & \cellcolor[rgb]{ 1,  .949,  .8}0.27 & \cellcolor[rgb]{ 1,  .949,  .8}0.27 & \cellcolor[rgb]{ 1,  .949,  .8}0.27 & \cellcolor[rgb]{ 1,  .949,  .8}0.27 & \cellcolor[rgb]{ 1,  .949,  .8}0.28 & \cellcolor[rgb]{ .973,  .816,  .8}0.29 & \cellcolor[rgb]{ .973,  .816,  .8}0.29 & \cellcolor[rgb]{ .973,  .816,  .8}0.29 & \cellcolor[rgb]{ .973,  .816,  .8}0.29 & \cellcolor[rgb]{ .973,  .816,  .8}0.29 \\
\hline \hline
\rowcolor[rgb]{ .851,  .851,  .851} $V_{nm}^{SH}$ & \cellcolor[rgb]{ .886,  .937,  .855}0.25 & \cellcolor[rgb]{ .886,  .937,  .855}0.25 & \cellcolor[rgb]{ .886,  .937,  .855}0.25 & \cellcolor[rgb]{ .886,  .937,  .855}0.25 & \cellcolor[rgb]{ .886,  .937,  .855}0.26 & \cellcolor[rgb]{ .867,  .922,  .969}0.34 & \cellcolor[rgb]{ .867,  .922,  .969}0.35 & \cellcolor[rgb]{ .867,  .922,  .969}0.35 & \cellcolor[rgb]{ .867,  .922,  .969}0.35 & \cellcolor[rgb]{ .867,  .922,  .969}0.36 & \cellcolor[rgb]{ 1,  .949,  .8}0.41 & \cellcolor[rgb]{ 1,  .949,  .8}0.41 & \cellcolor[rgb]{ 1,  .949,  .8}0.41 & \cellcolor[rgb]{ 1,  .949,  .8}0.41 & \cellcolor[rgb]{ 1,  .949,  .8}0.42 & \cellcolor[rgb]{ .973,  .816,  .8}0.54 & \cellcolor[rgb]{ .973,  .816,  .8}0.54 & \cellcolor[rgb]{ .973,  .816,  .8}0.54 & \cellcolor[rgb]{ .973,  .816,  .8}0.54 & \cellcolor[rgb]{ .973,  .816,  .8}0.55 \\
\hline
\rowcolor[rgb]{ .851,  .851,  .851} $V_{\delta_{nm}}^{SH}$ & \cellcolor[rgb]{ .886,  .937,  .855}0.34 & \cellcolor[rgb]{ .886,  .937,  .855}0.33 & \cellcolor[rgb]{ .886,  .937,  .855}0.34 & \cellcolor[rgb]{ .886,  .937,  .855}0.35 & \cellcolor[rgb]{ .886,  .937,  .855}0.35 & \cellcolor[rgb]{ .867,  .922,  .969}0.41 & \cellcolor[rgb]{ .867,  .922,  .969}0.41 & \cellcolor[rgb]{ .867,  .922,  .969}0.41 & \cellcolor[rgb]{ .867,  .922,  .969}0.41 & \cellcolor[rgb]{ .867,  .922,  .969}0.41 & \cellcolor[rgb]{ 1,  .949,  .8}0.47 & \cellcolor[rgb]{ 1,  .949,  .8}0.47 & \cellcolor[rgb]{ 1,  .949,  .8}0.47 & \cellcolor[rgb]{ 1,  .949,  .8}0.48 & \cellcolor[rgb]{ 1,  .949,  .8}0.48 & \cellcolor[rgb]{ .973,  .816,  .8}0.61 & \cellcolor[rgb]{ .973,  .816,  .8}0.61 & \cellcolor[rgb]{ .973,  .816,  .8}0.61 & \cellcolor[rgb]{ .973,  .816,  .8}0.66 & \cellcolor[rgb]{ .973,  .816,  .8}0.66 \\
\hline
\rowcolor[rgb]{ .851,  .851,  .851} $V_{\theta_{nm}}^{SH}$ & \cellcolor[rgb]{ .886,  .937,  .855}0.23 & \cellcolor[rgb]{ .886,  .937,  .855}0.22 & \cellcolor[rgb]{ .886,  .937,  .855}0.22 & \cellcolor[rgb]{ .886,  .937,  .855}0.22 & \cellcolor[rgb]{ .886,  .937,  .855}0.23 & \cellcolor[rgb]{ .867,  .922,  .969}0.37 & \cellcolor[rgb]{ .867,  .922,  .969}0.37 & \cellcolor[rgb]{ .867,  .922,  .969}0.37 & \cellcolor[rgb]{ .867,  .922,  .969}0.37 & \cellcolor[rgb]{ .867,  .922,  .969}0.38 & \cellcolor[rgb]{ 1,  .949,  .8}0.44 & \cellcolor[rgb]{ 1,  .949,  .8}0.44 & \cellcolor[rgb]{ 1,  .949,  .8}0.44 & \cellcolor[rgb]{ 1,  .949,  .8}0.44 & \cellcolor[rgb]{ 1,  .949,  .8}0.45 & \cellcolor[rgb]{ .973,  .816,  .8}0.55 & \cellcolor[rgb]{ .973,  .816,  .8}0.55 & \cellcolor[rgb]{ .973,  .816,  .8}0.55 & \cellcolor[rgb]{ .973,  .816,  .8}0.55 & \cellcolor[rgb]{ .973,  .816,  .8}0.56 \\
\hline
\rowcolor[rgb]{ .851,  .851,  .851} $V_{\alpha_{nm}}^{SH}$ & \cellcolor[rgb]{ .886,  .937,  .855}0.2 & \cellcolor[rgb]{ .886,  .937,  .855}0.2 & \cellcolor[rgb]{ .886,  .937,  .855}0.19 & \cellcolor[rgb]{ .886,  .937,  .855}0.2 & \cellcolor[rgb]{ .886,  .937,  .855}0.2 & \cellcolor[rgb]{ .867,  .922,  .969}0.34 & \cellcolor[rgb]{ .867,  .922,  .969}0.34 & \cellcolor[rgb]{ .867,  .922,  .969}0.34 & \cellcolor[rgb]{ .867,  .922,  .969}0.34 & \cellcolor[rgb]{ .867,  .922,  .969}0.36 & \cellcolor[rgb]{ 1,  .949,  .8}0.38 & \cellcolor[rgb]{ 1,  .949,  .8}0.38 & \cellcolor[rgb]{ 1,  .949,  .8}0.38 & \cellcolor[rgb]{ 1,  .949,  .8}0.38 & \cellcolor[rgb]{ 1,  .949,  .8}0.38 & \cellcolor[rgb]{ .973,  .816,  .8}0.5 & \cellcolor[rgb]{ .973,  .816,  .8}0.5 & \cellcolor[rgb]{ .973,  .816,  .8}0.5 & \cellcolor[rgb]{ .973,  .816,  .8}0.5 & \cellcolor[rgb]{ .973,  .816,  .8}0.51 \\
\hline
\rowcolor[rgb]{ .851,  .851,  .851} $V_{\beta_{nm}}^{SH}$ & \cellcolor[rgb]{ .886,  .937,  .855}0.17 & \cellcolor[rgb]{ .886,  .937,  .855}0.17 & \cellcolor[rgb]{ .886,  .937,  .855}0.16 & \cellcolor[rgb]{ .886,  .937,  .855}0.16 & \cellcolor[rgb]{ .886,  .937,  .855}0.17 & \cellcolor[rgb]{ .867,  .922,  .969}0.29 & \cellcolor[rgb]{ .867,  .922,  .969}0.29 & \cellcolor[rgb]{ .867,  .922,  .969}0.29 & \cellcolor[rgb]{ .867,  .922,  .969}0.29 & \cellcolor[rgb]{ .867,  .922,  .969}0.29 & \cellcolor[rgb]{ 1,  .949,  .8}0.33 & \cellcolor[rgb]{ 1,  .949,  .8}0.33 & \cellcolor[rgb]{ 1,  .949,  .8}0.33 & \cellcolor[rgb]{ 1,  .949,  .8}0.33 & \cellcolor[rgb]{ 1,  .949,  .8}0.34 & \cellcolor[rgb]{ .973,  .816,  .8}0.4 & \cellcolor[rgb]{ .973,  .816,  .8}0.4 & \cellcolor[rgb]{ .973,  .816,  .8}0.4 & \cellcolor[rgb]{ .973,  .816,  .8}0.4 & \cellcolor[rgb]{ .973,  .816,  .8}0.41 \\
\hline
\rowcolor[rgb]{ .851,  .851,  .851} $V_{\gamma_{nm}}^{SH}$ & \cellcolor[rgb]{ .886,  .937,  .855}0.09 & \cellcolor[rgb]{ .886,  .937,  .855}0.1 & \cellcolor[rgb]{ .886,  .937,  .855}0.1 & \cellcolor[rgb]{ .886,  .937,  .855}0.1 & \cellcolor[rgb]{ .886,  .937,  .855}0.1 & \cellcolor[rgb]{ .867,  .922,  .969}0.18 & \cellcolor[rgb]{ .867,  .922,  .969}0.18 & \cellcolor[rgb]{ .867,  .922,  .969}0.18 & \cellcolor[rgb]{ .867,  .922,  .969}0.18 & \cellcolor[rgb]{ .867,  .922,  .969}0.18 & \cellcolor[rgb]{ 1,  .949,  .8}0.28 & \cellcolor[rgb]{ 1,  .949,  .8}0.28 & \cellcolor[rgb]{ 1,  .949,  .8}0.28 & \cellcolor[rgb]{ 1,  .949,  .8}0.28 & \cellcolor[rgb]{ 1,  .949,  .8}0.3 & \cellcolor[rgb]{ .973,  .816,  .8}0.3 & \cellcolor[rgb]{ .973,  .816,  .8}0.3 & \cellcolor[rgb]{ .973,  .816,  .8}0.3 & \cellcolor[rgb]{ .973,  .816,  .8}0.3 & \cellcolor[rgb]{ .973,  .816,  .8}0.31 \\
\hline \hline
\rowcolor[rgb]{ .851,  .851,  .851} $V_{nm}^{H2}$ & \cellcolor[rgb]{ .886,  .937,  .855}0.25 & \cellcolor[rgb]{ .886,  .937,  .855}0.25 & \cellcolor[rgb]{ .886,  .937,  .855}0.26 & \cellcolor[rgb]{ .886,  .937,  .855}0.26 & \cellcolor[rgb]{ .886,  .937,  .855}0.26 & \cellcolor[rgb]{ .867,  .922,  .969}0.35 & \cellcolor[rgb]{ .867,  .922,  .969}0.35 & \cellcolor[rgb]{ .867,  .922,  .969}0.35 & \cellcolor[rgb]{ .867,  .922,  .969}0.35 & \cellcolor[rgb]{ .867,  .922,  .969}0.36 & \cellcolor[rgb]{ 1,  .949,  .8}0.42 & \cellcolor[rgb]{ 1,  .949,  .8}0.42 & \cellcolor[rgb]{ 1,  .949,  .8}0.42 & \cellcolor[rgb]{ 1,  .949,  .8}0.42 & \cellcolor[rgb]{ 1,  .949,  .8}0.43 & \cellcolor[rgb]{ .973,  .816,  .8}0.55 & \cellcolor[rgb]{ .973,  .816,  .8}0.55 & \cellcolor[rgb]{ .973,  .816,  .8}0.55 & \cellcolor[rgb]{ .973,  .816,  .8}0.55 & \cellcolor[rgb]{ .973,  .816,  .8}0.57 \\
\hline
\rowcolor[rgb]{ .851,  .851,  .851} $V_{\delta_{nm}}^{H2}$ & \cellcolor[rgb]{ .886,  .937,  .855}0.34 & \cellcolor[rgb]{ .886,  .937,  .855}0.33 & \cellcolor[rgb]{ .886,  .937,  .855}0.34 & \cellcolor[rgb]{ .886,  .937,  .855}0.34 & \cellcolor[rgb]{ .886,  .937,  .855}0.35 & \cellcolor[rgb]{ .867,  .922,  .969}0.41 & \cellcolor[rgb]{ .867,  .922,  .969}0.41 & \cellcolor[rgb]{ .867,  .922,  .969}0.41 & \cellcolor[rgb]{ .867,  .922,  .969}0.41 & \cellcolor[rgb]{ .867,  .922,  .969}0.41 & \cellcolor[rgb]{ 1,  .949,  .8}0.48 & \cellcolor[rgb]{ 1,  .949,  .8}0.48 & \cellcolor[rgb]{ 1,  .949,  .8}0.48 & \cellcolor[rgb]{ 1,  .949,  .8}0.48 & \cellcolor[rgb]{ 1,  .949,  .8}0.49 & \cellcolor[rgb]{ .973,  .816,  .8}0.62 & \cellcolor[rgb]{ .973,  .816,  .8}0.62 & \cellcolor[rgb]{ .973,  .816,  .8}0.62 & \cellcolor[rgb]{ .973,  .816,  .8}0.62 & \cellcolor[rgb]{ .973,  .816,  .8}0.65 \\
\hline
\rowcolor[rgb]{ .851,  .851,  .851} $V_{\theta_{nm}}^{H2}$ & \cellcolor[rgb]{ .886,  .937,  .855}0.25 & \cellcolor[rgb]{ .886,  .937,  .855}0.24 & \cellcolor[rgb]{ .886,  .937,  .855}0.24 & \cellcolor[rgb]{ .886,  .937,  .855}0.23 & \cellcolor[rgb]{ .886,  .937,  .855}0.23 & \cellcolor[rgb]{ .867,  .922,  .969}0.38 & \cellcolor[rgb]{ .867,  .922,  .969}0.38 & \cellcolor[rgb]{ .867,  .922,  .969}0.38 & \cellcolor[rgb]{ .867,  .922,  .969}0.39 & \cellcolor[rgb]{ .867,  .922,  .969}0.39 & \cellcolor[rgb]{ 1,  .949,  .8}0.44 & \cellcolor[rgb]{ 1,  .949,  .8}0.44 & \cellcolor[rgb]{ 1,  .949,  .8}0.44 & \cellcolor[rgb]{ 1,  .949,  .8}0.44 & \cellcolor[rgb]{ 1,  .949,  .8}0.45 & \cellcolor[rgb]{ .973,  .816,  .8}0.53 & \cellcolor[rgb]{ .973,  .816,  .8}0.53 & \cellcolor[rgb]{ .973,  .816,  .8}0.53 & \cellcolor[rgb]{ .973,  .816,  .8}0.53 & \cellcolor[rgb]{ .973,  .816,  .8}0.55 \\
\hline
\rowcolor[rgb]{ .851,  .851,  .851} $V_{\alpha_{nm}}^{H2}$ & \cellcolor[rgb]{ .886,  .937,  .855}0.22 & \cellcolor[rgb]{ .886,  .937,  .855}0.2 & \cellcolor[rgb]{ .886,  .937,  .855}0.2 & \cellcolor[rgb]{ .886,  .937,  .855}0.2 & \cellcolor[rgb]{ .886,  .937,  .855}0.22 & \cellcolor[rgb]{ .867,  .922,  .969}0.36 & \cellcolor[rgb]{ .867,  .922,  .969}0.36 & \cellcolor[rgb]{ .867,  .922,  .969}0.36 & \cellcolor[rgb]{ .867,  .922,  .969}0.36 & \cellcolor[rgb]{ .867,  .922,  .969}0.36 & \cellcolor[rgb]{ 1,  .949,  .8}0.38 & \cellcolor[rgb]{ 1,  .949,  .8}0.38 & \cellcolor[rgb]{ 1,  .949,  .8}0.38 & \cellcolor[rgb]{ 1,  .949,  .8}0.38 & \cellcolor[rgb]{ 1,  .949,  .8}0.39 & \cellcolor[rgb]{ .973,  .816,  .8}0.51 & \cellcolor[rgb]{ .973,  .816,  .8}0.51 & \cellcolor[rgb]{ .973,  .816,  .8}0.51 & \cellcolor[rgb]{ .973,  .816,  .8}0.51 & \cellcolor[rgb]{ .973,  .816,  .8}0.51 \\
\hline
\rowcolor[rgb]{ .851,  .851,  .851} $V_{\beta_{nm}}^{H2}$ & \cellcolor[rgb]{ .886,  .937,  .855}0.18 & \cellcolor[rgb]{ .886,  .937,  .855}0.18 & \cellcolor[rgb]{ .886,  .937,  .855}0.18 & \cellcolor[rgb]{ .886,  .937,  .855}0.18 & \cellcolor[rgb]{ .886,  .937,  .855}0.18 & \cellcolor[rgb]{ .867,  .922,  .969}0.28 & \cellcolor[rgb]{ .867,  .922,  .969}0.28 & \cellcolor[rgb]{ .867,  .922,  .969}0.28 & \cellcolor[rgb]{ .867,  .922,  .969}0.28 & \cellcolor[rgb]{ .867,  .922,  .969}0.29 & \cellcolor[rgb]{ 1,  .949,  .8}0.33 & \cellcolor[rgb]{ 1,  .949,  .8}0.33 & \cellcolor[rgb]{ 1,  .949,  .8}0.33 & \cellcolor[rgb]{ 1,  .949,  .8}0.33 & \cellcolor[rgb]{ 1,  .949,  .8}0.34 & \cellcolor[rgb]{ .973,  .816,  .8}0.39 & \cellcolor[rgb]{ .973,  .816,  .8}0.39 & \cellcolor[rgb]{ .973,  .816,  .8}0.39 & \cellcolor[rgb]{ .973,  .816,  .8}0.39 & \cellcolor[rgb]{ .973,  .816,  .8}0.4 \\
\hline
\rowcolor[rgb]{ .851,  .851,  .851} $V_{\gamma_{nm}}^{H2}$ & \cellcolor[rgb]{ .886,  .937,  .855}0.11 & \cellcolor[rgb]{ .886,  .937,  .855}0.11 & \cellcolor[rgb]{ .886,  .937,  .855}0.11 & \cellcolor[rgb]{ .886,  .937,  .855}0.11 & \cellcolor[rgb]{ .886,  .937,  .855}0.11 & \cellcolor[rgb]{ .867,  .922,  .969}0.16 & \cellcolor[rgb]{ .867,  .922,  .969}0.16 & \cellcolor[rgb]{ .867,  .922,  .969}0.16 & \cellcolor[rgb]{ .867,  .922,  .969}0.16 & \cellcolor[rgb]{ .867,  .922,  .969}0.17 & \cellcolor[rgb]{ 1,  .949,  .8}0.2 & \cellcolor[rgb]{ 1,  .949,  .8}0.2 & \cellcolor[rgb]{ 1,  .949,  .8}0.2 & \cellcolor[rgb]{ 1,  .949,  .8}0.2 & \cellcolor[rgb]{ 1,  .949,  .8}0.2 & \cellcolor[rgb]{ .973,  .816,  .8}0.19 & \cellcolor[rgb]{ .973,  .816,  .8}0.19 & \cellcolor[rgb]{ .973,  .816,  .8}0.19 & \cellcolor[rgb]{ .973,  .816,  .8}0.19 & \cellcolor[rgb]{ .973,  .816,  .8}0.19 \\
\hline \hline
\rowcolor[rgb]{ .851,  .851,  .851} $V_{com}$ & \cellcolor[rgb]{ .886,  .937,  .855}0.36 & \cellcolor[rgb]{ .886,  .937,  .855}0.36 & \cellcolor[rgb]{ .886,  .937,  .855}0.36 & \cellcolor[rgb]{ .886,  .937,  .855}0.36 & \cellcolor[rgb]{ .886,  .937,  .855}0.37 & \cellcolor[rgb]{ .867,  .922,  .969}0.43 & \cellcolor[rgb]{ .867,  .922,  .969}0.43 & \cellcolor[rgb]{ .867,  .922,  .969}0.44 & \cellcolor[rgb]{ .867,  .922,  .969}0.44 & \cellcolor[rgb]{ .867,  .922,  .969}0.44 & \cellcolor[rgb]{ 1,  .949,  .8}0.5 & \cellcolor[rgb]{ 1,  .949,  .8}0.5 & \cellcolor[rgb]{ 1,  .949,  .8}0.5 & \cellcolor[rgb]{ 1,  .949,  .8}0.51 & \cellcolor[rgb]{ 1,  .949,  .8}0.52 & \cellcolor[rgb]{ .973,  .816,  .8}0.67 & \cellcolor[rgb]{ .973,  .816,  .8}0.67 & \cellcolor[rgb]{ .973,  .816,  .8}0.67 & \cellcolor[rgb]{ .973,  .816,  .8}0.68 & \cellcolor[rgb]{ .973,  .816,  .8}0.70 \\
\hline
\end{tabular}}
\label{tab:Tabper}
\scriptsize{
\textcolor[rgb]{ .886,  .937,  .855}{$\blacksquare$}: $320$ ms window, \textcolor[rgb]{ .867,  .922,  .969}{$\blacksquare$}: $800$ ms window, \textcolor[rgb]{ 1,  .949,  .8}{$\blacksquare$}: $1200$ ms window , \textcolor[rgb]{ .973,  .816,  .8}{$\blacksquare$}: $1600$ ms window}; Note: $V_{com}: [V_{\delta};V_{\delta_{nm}}^{SH};V_{\delta_{nm}}^{H2}]$
\end{table*}
\item Dense layers: Three dense layers with $8$ units each are used after the flatten layer. In this work, the swish activation function is used in these layers, interpreted as:
\begin{equation}
f(x)=x\ . \ swish(x)
\end{equation}
\item Output layer: The final layer is a dense layer for regression that maps the output of flatten layer to the predicted trajectory with dimension $N \times 1$, as shown in Fig. \ref{fig:figmethod}. Dense layer implements the element-wise dot product between the input and the kernel. The linear activation function is used in this layer, given by:
\begin{equation}
f(x)=x
\end{equation}
\end{itemize}

The aforementioned layers and hyper-parameters are used to create the proposed network. For training, $80\%$ of EEG signals with different durations/window lengths are taken from the recorded database. The rest $20\%$ of the data is divided into $10\%$ test and $10\%$ validation data. The information about optimal training hyper-parameter selection and their values is provided in the next section. The proposed network is built using the TensorFlow deep learning framework with version 2.2.1 as the backend in Python. This work uses data augmentation to increase the number of training examples in the data to avoid over-fitting. It makes the proposed network more robust by creating new and different training examples by which it can learn the alterations in the real world. For this purpose, random flipping and rolling operations are used in the Python Keras framework. 

\section{Results and Discussion}
In this Section, the performance evaluation of the proposed BiCurNet on the recorded EEG signals is presented w.r.t. different parameters. Elaborated interpretations of the results are also presented for the proposed network.
 
\subsection{Hyper-parameters selection}
Various parameters used for training the proposed network are presented herein. For assessing the decoding performance of the proposed network, $10\%$ of the EEG signals from the recorded database are used for testing. The data from each subject is used for training, testing, and validation for subject-dependent case. The leave-one-subject-out scheme is adopted for subject-independent performance analysis. The network is trained using a batch size of $15$, epochs as $100$, and an Adam optimizer with a learning rate of $0.001$. Mean square error (MSE) is used as the loss function for regression. Table \ref{Tabhyper} presents the training hyper-parameters selected using the KerasTuner framework in Python. It is an optimization framework for tuning the hyper-parameters that uses search-and-selection-based criteria. The final corresponding selected set of optimal hyper-parameters is listed in the table. {\color{red}The system utilized for conducting the experiment has an Intel Core i7-8550U CPU with 8 GB RAM. The deep learning models are employed using the Python Tensorflow 2 platform. The running time for the trained model on EEG sub-bands test data with window sizes of 320 ms, 800 ms, 1200 ms, and 1600 ms is approximately 11 ms, 18 ms, 19 ms, and 19 ms, respectively. However, the running time for harmonics-domain features of EEG sub-bands is approximately 1 ms, 8 ms, 12 ms, and 15 ms for 320 ms, 800 ms, 1200 ms, and 1600 ms window sizes, respectively.}

\subsection{Regression metric}
This work uses time-lagged and windowed EEG signals to estimate the motion trajectory in advance. In particular, the EEG data preceding the motion by different time lags ($8$-$240$ ms) is used to train, test, and validate the proposed network. Additionally, the performance is evaluated with varying EEG window sizes ($320$-$1600$ ms). A $95\%$ overlap between adjacent windows is considered.
\subsubsection{Pearson Correlation Coefficient}
Pearson correlation coefficient (PCC) is utilized for analyzing the performance of the proposed network w.r.t. upper limb motion trajectory estimation. PCC between true/measured ($A$) and predicted/estimated ($P$) trajectory signal with $N$ samples is given as
\begin{equation}
\Pi(A,P)=\frac{1}{N-1} \sum_{i=1}^N\left(\frac{A_i-m_A}{\sigma_A}\right)\left(\frac{P_i-m_P}{\sigma_P}\right)
\end{equation}
where $m$ is the mean and $\sigma$ denotes standard deviation. The normalized covariance measure assumes a value between -1 and 1.

{\color{red}\subsubsection{Root Mean Squared Error}
Root Mean Square Error (RMSE) is a widely used metric for evaluating the accuracy of predictive models. It quantifies the average discrepancy between predicted values \( \hat{y}_i \) and actual values \( y_i \) within a dataset, thus providing a measure of the model's performance. RMSE is computed as the square root of the mean of the squared differences between the predicted and actual values for \( i = 1, 2, \ldots, n \), where \( n \) represents the total number of data points. The formula for RMSE is as follows:
\[
RMSE = \sqrt{\frac{1}{n} \sum_{i=1}^{n} (\hat{y}_i - y_i)^2}
\]
In this equation, \( \hat{y}_i \) represents the predicted value for the \(i\)-th data point, and \( y_i \) denotes the corresponding actual value. A lower RMSE value indicates better alignment between the model's predictions and the actual data.
}

\subsection{Subject-dependent performance analysis}
 The proposed model is trained and tested for each subject separately for subject-dependent (SD) performance analysis. The PCC values averaged across all the trials and subjects are presented in Table \ref{tab:Tabper} with varying time lags, window sizes, and EEG features. The EEG bands are considered in spatial ($V$), spherical harmonics ($V_{\delta_{nm}}$), and head harmonics domains ($V_{\delta_{nm}}^{H2}$). It may be noted that the transformed domain ($V_{\delta_{nm}}$ and $V_{\delta_{nm}}^{H2}$) features give PCC similar to spatial domain counterparts with reduced computational cost, as detailed in Section \ref{dwt-sh}.
 Additionally, the $\delta$ band gives higher PCC values while 
 $\gamma$ band has the lowest PCC. This indicates the pertinence of the low-frequency $\delta$ band for motion trajectory decoding using EEG.
The best correlation is observed when $V_\delta$, $V_{\delta_{nm}}^{SH}$, and $V_{\delta_{nm}}^{H2}$ are combined, i.e. $V_{com}$. The highest correlation achieved is $0.7$ with $240$ ms advanced EEG window of $1600$ ms. Also, the lowest RMSE value of $0.11$ is achieved with a lag of $240$ ms and $1600$ ms window size. This demonstrates the feasibility of early estimation of the motion trajectory by using the proposed network.

\subsection{Subject-independent performance analysis}
\begin{figure}[h]
\includegraphics[width=7.5cm, height=4.5cm]{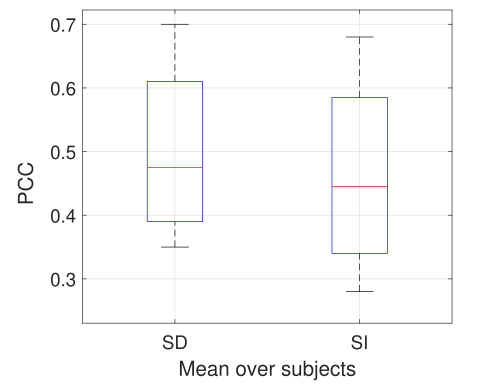}
\caption{Average PCC values w.r.t. subject dependent (SD) and subject-independent (SI) training of the proposed network at different window sizes ($320$ ms to $1600$ ms).}
\label{fig:figsdsi}
\end{figure}

 To further explore the adaptability of the proposed network, subject-independent (SI) analysis is presented herein using a leave-one-out scheme. A simultaneous comparison of SI/SD cases on PCC is presented in Fig. \ref{fig:figsdsi}. The PCC values are averaged over all subjects and lags. A slight decrease in PCC value may be noted in the SI case. However, it remains within $\pm 0.05$, indicating the proposed network's robustness against the subject-variability. The proposed BicurNet decoder achieved the highest PCC value of $0.62$ and the lowest RMSE value of $0.17$ using $V_{com}$ with a $1600$ ms window and $240$ ms lag.

\begin{figure}[!h]
\includegraphics[width=7.5cm,height=4.56cm]{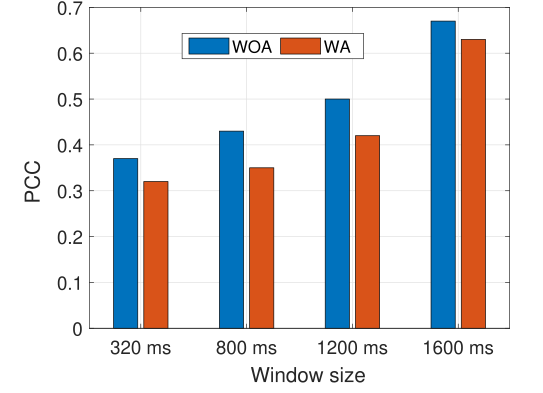}
\caption{Subject-dependent average PCC values utilizing with and without artifactual EEG data for different window sizes.}
\label{fig:fignoiserobst}
\end{figure}
\subsection{Robustness analysis} 
The robustness of the proposed network is analyzed herein using artifactual EEG signals. In particular, the pre-processing did not include ICA decomposition-based artifact removal. The proposed network is trained and tested using such signals. Mean PCC values obtained using without artifact (WOA) and with artifact (WA) EEG signal are presented in Fig. \ref{fig:fignoiserobst}. A small decrease of $0.06$ in the PCC values may be observed with an artifact case that indicates the robustness of the proposed model.

\subsection{Trajectory estimation curves}
\begin{figure}[!t]
\includegraphics[width=0.4\textwidth]{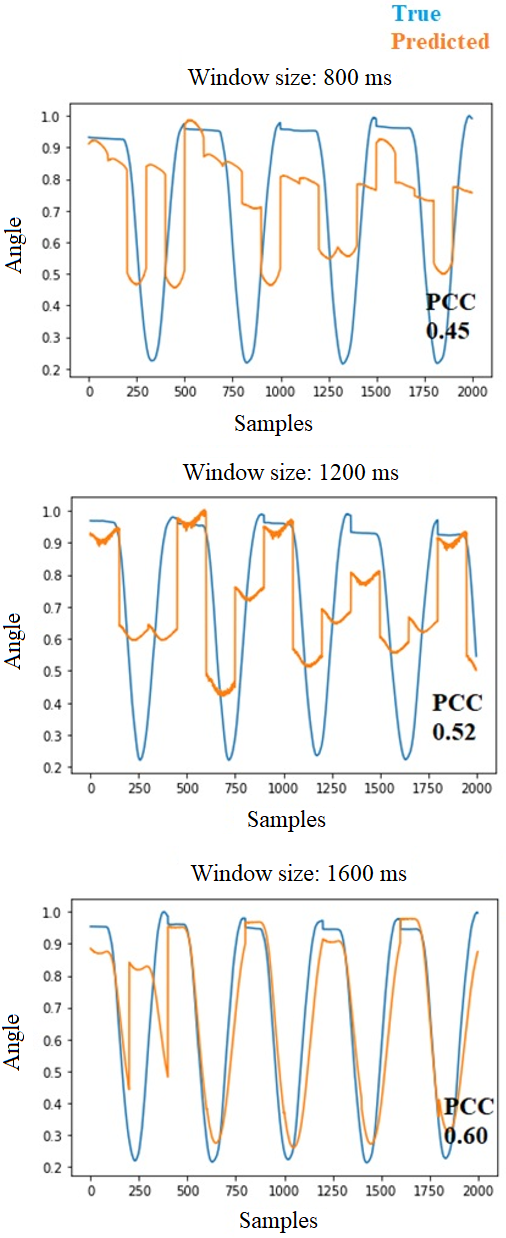}
\caption{Actual and predicted trajectories of subject 1 (Early prediction, before 40 ms).}
\label{fig:figtrajectory}
\end{figure}

The proposed BiCurNet model is additionally evaluated herein using actual motion trajectories. Fig. \ref{fig:figtrajectory} illustrates the estimated and actual trajectories for subject I with window size varying between $800$-$1600$ ms. 95\% overlap is considered between two adjacent windows. It may be observed from the figure that there is a considerable improvement in correlation when window size is increased. This results in a trajectory closer to the ground truth. The ability of the proposed network to follow the trajectory pattern for all windows indicates the learning capability of the network.

\begin{figure*}[t]
	\centering
        \subfigure[]
	{\includegraphics[width=0.45\textwidth]{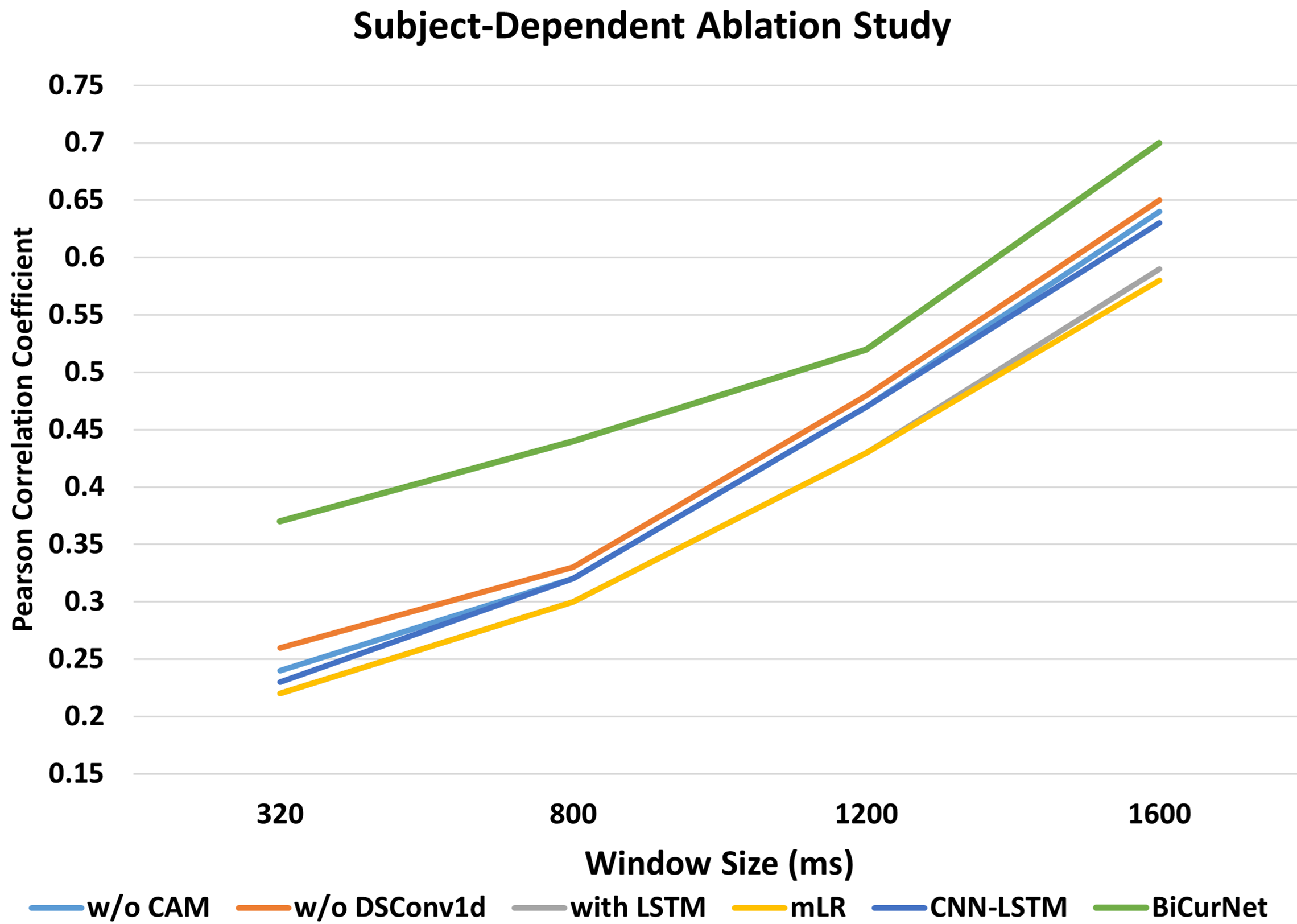}}
        \subfigure[]
	{\includegraphics[width=0.45\textwidth]{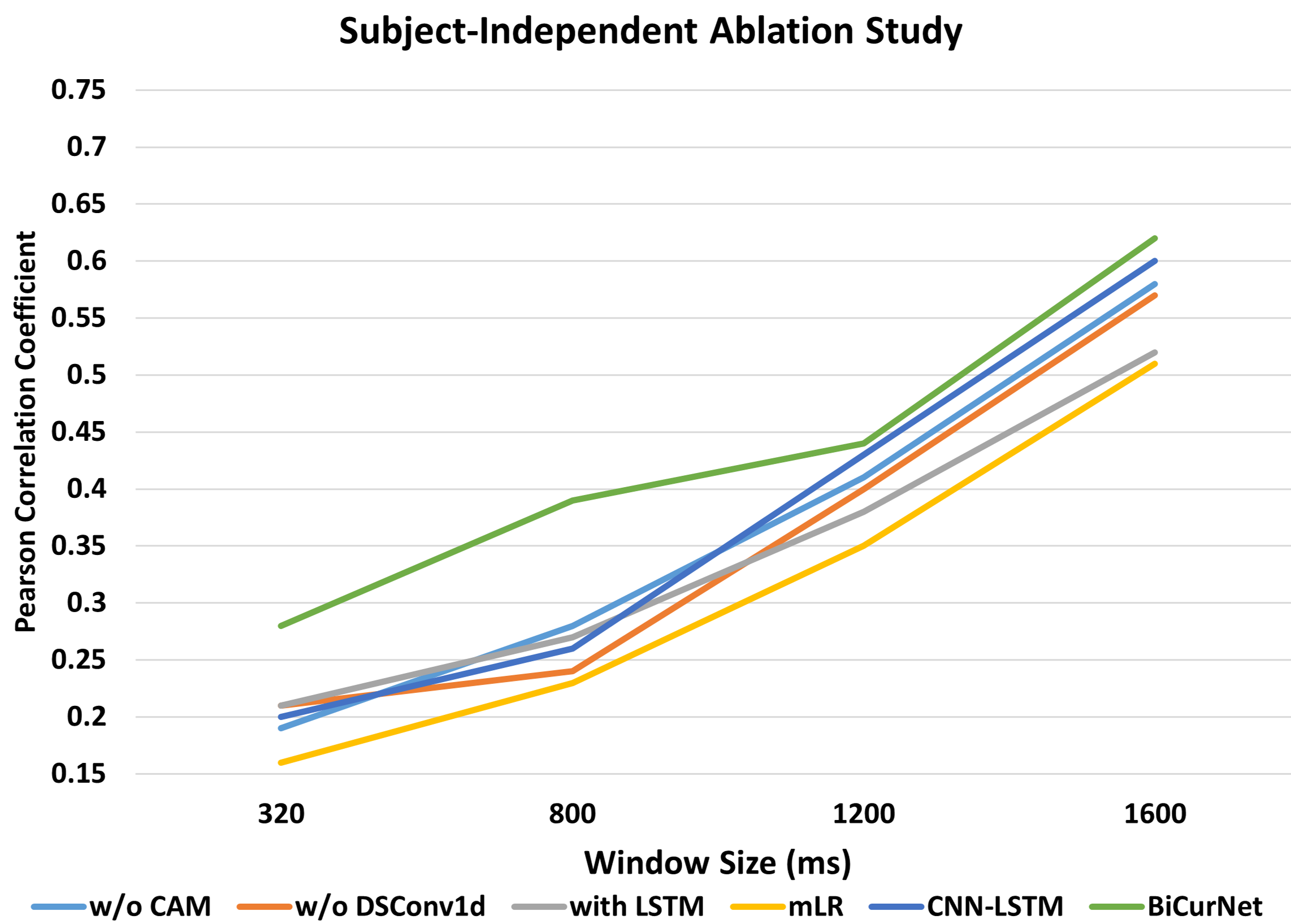}}
	\caption{Network ablation and comparison for (a) Subject-dependent and (b) Subject-Independent settings.}
	\label{ablation} 
\end{figure*}

{\color{red}\subsection{Ablation Study}
In this Section, an ablation study of the BiCurNet model is presented. The ablation study is performed to analyze the influence of the DWSConv1D and custom attention module (CAM) on the performance of the proposed model. Additionally, a comparison with the multi-linear regression (mLR) model is performed. The study is performed for both subject-dependent and subject-independent setups. A time lag of 240 ms and a window size of up to 1600 ms are utilized for the analysis. A combined feature set of $V_\delta$, $V_{\delta_{nm}}^{SH}$, and $V_{\delta_{nm}}^{H2}$, i.e. $V_{com}$ is considered for model training and testing.

The PCC analysis of the ablation study for subject-dependent and subject-independent is shown in Fig. \ref{ablation}(a) and \ref{ablation}(b), respectively. The proposed BiCurNet model outperforms the traditional mLR model in both cases. The performance of the proposed model without CAM and DSConv1D layers also degraded, showing the importance of these layers for efficient trajectory decoding.
}
\section{Conclusion}
 A deep learning-based paradigm for early estimation of upper limb motion trajectory using EEG signal is proposed in this work. The EEG is collected while performing biceps curl movement. The proposed BiCurNet model uses a lightweight architecture with depth-wise separable convolution layers and a customized attention module. The input features to the model are taken in computationally more efficient spherical and head harmonics domains in addition to spatio-temporal data. The extensive performance evaluation of the proposed network on in-house recorded EEG signals demonstrates its effectiveness in early estimation. Performance evaluation includes subject (in)dependent study. The noise awareness of the proposed network is also demonstrated by using artifactual EEG signals for training. The robustness of the proposed network is demonstrated by using the artifactual EEG signals for training. The proposed network is computationally efficient and noise-aware, making it suitable for real-time BCI applications. Real-time implementation of the proposed network for an exosuit control is currently being explored.
 

\section*{Acknowledgment}
This research work was supported in part by the DRDO - JATC project with project number RP04191G.

\balance
\bibliographystyle{IEEEtran}

\bibliography{MTP_EEG_Paper_rev}

\end{document}